\documentclass[a4paper,12pt,preprintnumbers]{revtex4}

\usepackage{amsmath,amssymb,amsfonts}
\usepackage{hyperref}
\usepackage{graphicx}
\usepackage{color}
\usepackage[utf8]{inputenc}

\newcommand{\desy}{DESY Theory Group, Notkestr.~85, D-22607 Hamburg, Germany}
\newcommand{\pavia}{INFN, Sezione di Pavia, Via A. Bassi 6, 27100 Pavia, Italy}
\newcommand{\naples}{Dipartimento di Fisica, Universit\`a di Napoli ``Federico II'' and INFN,
sezione di Napoli, via Cintia, 80126 Napoli, Italy}

\begin{document}
\preprint{DESY 15-113}
\preprint{FNT/2015/01}
\title{Automation of electroweak corrections for LHC processes}
\author{Mauro~Chiesa}
\affiliation{\pavia}
\author{Nicolas~Greiner}
\affiliation{\desy}
\author{Francesco Tramontano}
\affiliation{\naples}
\begin{abstract}
\noindent
For the Run 2 of the LHC next-to-leading order electroweak corrections will play an important role. 
Even though they are typically moderate at the level of total cross sections they can lead to substantial deviations
in the shapes of distributions. In particular for new physics searches but also for a precise determination
of Standard Model observables their inclusion in the 
theoretical predictions is mandatory for a reliable estimation of the Standard Model contribution.
In this article we review the status and recent developments 
in electroweak calculations and their automation for LHC processes. We discuss general issues and properties of NLO 
electroweak corrections and present some examples, including the full calculation of the NLO corrections to  
the production of a $W$-boson in association with two jets 
computed using {\sc{GoSam}} interfaced to {\sc{MadDipole}}.
~\\
\end{abstract}

\maketitle

\tableofcontents
\section{Introduction}
The Large Hadron Collider (LHC) was designed to provide a deeper insight into
the microscopic world and increase our knowledge about the fundamental
interactions occurring in nature.

The recent discovery of the Higgs boson~\cite{Chatrchyan:2012ufa,Aad:2012tfa} 
supports the predictions made by the Standard Model of particle physics~
\cite{Englert:1964et,Higgs:1964ia,Higgs:1964pj,Guralnik:1964eu,Kibble:1967sv}.
It is not yet clear though, whether the discovered boson is exactly 
the one predicted by the Standard Model, or whether it originates from supersymmetric 
theories, or whether it is something completely different mimicking 
the role and the properties of a Higgs boson. 

There is the hope that the run~2 at the LHC will show not only more 
phenomenological properties of the Higgs particle, but also additional new 
physics beyond the Standard Model (BSM). 
As in principle it is not known how new physics will manifest itself, 
one has to look for even small deviations from the Standard Model 
predictions. 
Completely different new physics scenarios can lead 
to very similar predictions, which makes it difficult to distinguish 
among the models. 
Besides precise experimental measurements this requires precise 
theoretical predictions.

At high energy hadron colliders, final states typically involve hundreds 
of particles (hadrons, leptons, neutrinos and photons). General purpose Monte Carlo (MC)  
event generators like 
{\sc{HERWIG}}~\cite{Corcella:2000bw}, 
{\sc{HERWIG++}}~\cite{Bahr:2008pv},  
{\sc{PYTHIA}}~6~\cite{Sjostrand:2006za},
{\sc{PYTHIA}}~8~\cite{Sjostrand:2007gs} and 
{\sc{SHERPA}}~\cite{Gleisberg:2008ta} 
can provide a fully exclusive description of such complicated 
reactions. These tools perform several tasks ranging from the computation 
of hard scattering processes to the simulation of multiple soft and/or 
collinear QCD and QED radiation 
%off the external particles of the hard scattering 
by means of parton shower algorithms. MC event generators also provide 
a description of the non-perturbative effects like hadronization and formation 
of the underlying event. General purpose MC generators play a crucial role 
in collider physics and indeed there is no experimental analysis or realistic 
phenomenological study which does not rely at least in part on these codes.

One of the drawbacks of general purpose MC event generators is the fact that 
hard scattering processes (and, as a consequence, overall normalizations) are 
computed at tree-level accuracy, while 
precise theoretical predictions for several processes of phenomenological 
interest  require at least the calculation of next-to-leading order (NLO) 
contributions in perturbation theory. 
Typically, at hadron colliders the largest contributions come from QCD corrections. 
In the last recent years, lots of progress has been made in calculating NLO QCD corrections: 
new algorithms, such as the ones based on the OPP 
method~\cite{Ossola:2006us,delAguila:2004nf,Mastrolia:2010nb} 
or on the generalized unitarity techniques~\cite{Bern:1994zx,Ellis:2007br,Ellis:2008ir},  
have been proposed to compute one loop virtual amplitudes numerically 
or semi-analytically, \emph{i.e.} bypassing the explicit calculation of one loop Feynman diagrams 
from the Feynman rules of the theory. In particular, a lot of effort
has been put into the automation of such intricate and time consuming calculations and 
several tools have been developed either based on generalized unitarity methods, 
such as {\sc{BlackHat}}~\cite{Berger:2008sj} and {\sc{Njet}}~\cite{Badger:2012pg}, 
or on OPP-inspired techniques, such as 
{\sc{GoSam}}~\cite{Cullen:2011ac,Cullen:2014yla}, 
{\sc{HELAC-NLO}}~\cite{Bevilacqua:2011xh}, 
{\sc{MadLoop}}~\cite{Hirschi:2011pa}/ {\sc{MadGraph}}~\cite{Alwall:2011uj,Alwall:2014hca}. 
In codes like {\sc{OpenLoops}}~\cite{Cascioli:2011va} and 
{\sc{RECOLA}}~\cite{Actis:2012qn} one loop amplitudes are 
computed numerically from tree level amplitudes by means of recursion relations 
applied to Feynman diagrams and off-shell currents, respectively. 
The fact that calculations were not  carried out any more on a process-by-process 
basis but obtained by automating the underlying building blocks and interfacing them 
in a standardized way, played a major role in what has become known  as the NLO revolution.

The NLO revolution of QCD corrections set a new standard for the data/theory comparison 
at run I of the LHC. While, at the TEVATRON, data were typically compared to the predictions  
of LO multi-parton event generators such as {\sc{ALPGEN}}~\cite{Mangano:2002ea}, 
{\sc{SHERPA}} and {\sc{MadGraph}} 
that were used to generate samples with different hard parton multiplicities merged 
to parton showers in the CKKW~\cite{Catani:2001cc,Krauss:2002up} 
or MLM~\cite{Mangano:2006rw} framework, the standard theoretical benchmark 
at the run I of the LHC for most of the analyses are NLO QCD computations consistently 
matched with parton showers within the 
POWHEG~\cite{Nason:2004rx,Frixione:2007vw,Alioli:2010xd} or the 
MC@NLO~\cite{Frixione:2002ik} framework. Methods 
like FxFx~\cite{Frederix:2012ps}, MEPS@NLO~\cite{Gehrmann:2012yg,Hoeche:2012yf}, 
MiNLO~\cite{Hamilton:2012np,Hamilton:2012rf}  
and UNLOPS~\cite{Lonnblad:2012ix} were also developed to merge parton 
showers and NLO samples with different parton multiplicity. Both matched 
and merged calculations rely on MC event generators like {\sc{HERWIG}}, 
{\sc{PYTHIA}} and {\sc{SHERPA}} for what concerns showering and non perturbative 
effects (for the description of multiple photon radiation off final state leptons 
the code {\sc{PHOTOS}}~\cite{Golonka:2005pn} is also widely used). 
Needless to say, this NLO revolution pushed the frontier of fixed order QCD corrections 
to the NNLO level of accuracy.

A lot of work has been done in the field of automation of one loop QCD corrections. 
However, as we will point out in the next paragraphs, in view of the 14 TeV run of the 
LHC there are the physical motivations to extend the technical know-how of
these automated approaches also to the context of electroweak corrections. Because 
of this point, we decided to focus the present review on the practical 
issues that one has to face in order to build a tool for the automated computation of 
NLO electroweak corrections to LHC processes starting from already available codes 
developed for the computation of the different building blocks of an NLO calculation. 
In particular, we will show how this task can be accomplished in
{\sc{GoSam}}$+${\sc{MadDipole}}~\cite{Frederix:2008hu,Frederix:2010cj,Gehrmann:2010ry} framework 
for the computation of the relevant amplitudes and the subtraction of 
infrared singularities respectively,
once the effect of electroweak renormalization has been properly 
taken into account.

The paper is organized as follows. In subsection~\ref{subsect:introphymot} we give 
a short overview about the phenomenological motivations for the calculation of NLO 
electroweak corrections at high energy hadron colliders. In this context, 
in subsection~\ref{subsect:introautomation} we explain the main reasons why 
the automation of $\mathcal{O}( \alpha )$ corrections will play a crucial role 
at the 14~TeV run of the LHC and, of course, at future colliders. We summarize 
the state of the art for the automation of NLO electroweak corrections, 
describing very briefly the existing tools and the phenomenological studies 
obtained with such codes. 
As stated above, with the present review we want to show how the automation of NLO electroweak 
corrections can be obtained starting from already existing tools: in particular 
we work in the {\sc{GoSam}}$+${\sc{MadDipole}} framework. 
In section~\ref{virtual} 
we discuss some of the issues related to the computation of virtual one loop 
electroweak corrections, such as the UV and IR regularization scheme dependence, 
the inclusion of electroweak renormalization and the subtraction of infrared 
singularities. As codes like {\sc{GoSam}} work in dimensional reduction~(DRED), 
while electroweak renormalization counterterms and infrared subtraction 
terms are usually derived in conventional dimensional regularization~(CDR), 
we give the transition rules from CDR to DRED regularization schemes for both 
renormalization counterterms and infrared subtraction terms. We show how these 
scheme conversion works with a few simple examples, checking explicitly the regularization 
scheme independence of the sum of virtual one loop corrections, renormalization 
counterterms and integrated dipoles. Real radiation effects, and in particular the 
treatment of photon radiation off massless fermions, are described in 
section~\ref{real}. As a realistic example of the technical issues described in 
sections~\ref{virtual} and~\ref{real}, we describe in detail a  
calculation of the one loop electroweak corrections to the process $pp \to W^+ jj$ 
in the {\sc{GoSam}}$+${\sc{MadDipole}} framework in section~\ref{wjj}. We show the results of some 
internal consistency checks concerning poles cancellation and we briefly discuss  
the phenomenological impact of the electroweak corrections to the $W$ production 
in association with $2$~jets.
   
\subsection{Physical motivations}
\label{subsect:introphymot}
In the next run of the LHC, with higher center of mass energy and also higher luminosity, 
electroweak radiative corrections will play a more and more important role.

First of all, the high luminosity will allow to perform electroweak precision 
measurements. This is for example the case for the $W$ 
boson mass measurements in the charged Drell-Yan process, where $M_W$ is 
determined indirectly from a template fit of the distributions of transverse mass 
$M_T=\sqrt{2 p_T^l p_T^{\nu} (1 - \rm{cos} \Delta \phi_{l \nu})}$ 
of the lepton neutrino system, of the the lepton $p_T$ and of the missing transverse energy. 
It is well known from the TEVATRON experience that a precise knowledge of the electroweak 
radiative corrections to the Drell-Yan is 
mandatory, as they alter the shapes of the above-mentioned distributions and thus affect the 
$W$ mass determination. This is the reason why different groups spent a lot of effort to 
compute the one loop electroweak corrections to the Drell-Yan and several simulation 
codes have been provided, such 
as {\sc{HORACE}}~\cite{CarloniCalame:2006zq,CarloniCalame:2007cd},
{\sc{RADY}}~\cite{Dittmaier:2001ay,Brensing:2007qm,Dittmaier:2009cr}, 
{\sc{SANC}}~\cite{Arbuzov:2005dd,Arbuzov:2007db}, 
{\sc{WGRAD}}~\cite{Baur:2004ig}, {\sc{WINHAC}} \cite{Placzek:2003zg,Bardin:2008fn} and 
{\sc{ZGRAD}}~\cite{Baur:2001ze}. All of these codes 
have LO QCD accuracy and NLO EW accuracy. In particular {\sc{SANC}} also include NLO QCD 
corrections, while {\sc{HORACE}} includes the effect of all order photon radiation properly 
matched to the $\mathcal{O}( \alpha )$ corrections. NLO EW corrections are added to 
the $\mathcal{O}( \alpha_S^2 )$ ones in the {\sc{FEWZ}} code~\cite{Li:2012wna}, while factorized NLO EW and NLO QCD 
corrections to the single $W$ and $Z$ production matched with QED and QCD parton shower 
have been implemented in the {\sc{POWHEG-BOX}} Monte Carlo event generator
~\cite{Barze:2012tt,Bernaciak:2012hj,Barze':2013yca}.

The higher luminosity will also improve the experimental accuracy for several other processes 
besides the Drell-Yan process. In order to match the foreseen experimental precision, theoretical predictions  
for many processes should include at least the effect of $\mathcal{O}( \alpha )$ corrections, as 
can be seen for example by looking at the Les Houches wish-list in Tables~1-3 
of Ref.~\cite{Butterworth:2014efa}. It is interesting 
to noting how the Les Houches wish-list also points out that, in order to fully exploit 
the physics potential of the LHC, fixed order EW computations should be properly matched 
with parton showers. At present, the issue of matching NLO EW corrections with parton showers 
has only been addressed for Drell-Yan in Refs.~\cite{Barze:2012tt,Barze':2013yca} 
and very recently for Higgs decay into four leptons in Ref.~\cite{Boselli:2015aha}.

%%%pdf
In the context of the high precision determination of the Standard Model parameters such as, 
for example, the $W$ boson mass, the theoretical uncertainties related to parton distribution 
functions (PDFs) will eventually become a limitation. Recently new PDF 
sets~\cite{Alekhin:2013nda,Ball:2014uwa,Harland-Lang:2014zoa,Dulat:2015mca} 
have been released 
which include the constraints coming from LHC run I data. PDFs are obtained by fitting 
experimental data with theoretical computations that are performed at the highest 
available perturbative order which at present is typically NNLO QCD. In Ref.~\cite{Ball:2014uwa} 
NLO electroweak corrections have been considered in the fit of Drell-Yan data~\cite{Boughezal:2013cwa}: the systematic 
inclusion of one loop EW corrections in PDF fits will move in the direction of reducing 
further the theoretical uncertainty related to PDFs. Besides the impact of EW corrections 
on PDF fits, electroweak corrections also affect the description of PDF evolution: QED corrections, 
for instance, can be included in the evolution equations, leading to the definition of 
the photon PDF~\cite{Martin:2004dh,Ball:2013hta,Schmidt:2015zda,Bertone:2013vaa,Carrazza:2014gfa}. 
In turn, the knowledge of the photon PDF allows to compute the so-called photon induced 
processes (i.e. partonic channels with photons in the initial state) which should be 
consistently included in the computation of $\mathcal{O}(\alpha)$ corrections\footnote{In 
Ref.~\cite{Bertone:2015lqa} the lepton PDF has also been determined. It may be relevant, for 
example, in the computation of EW corrections to photon initiated processes.}. 
At variance with the case of QED, the effect of weak corrections to PDF 
evolution~\cite{Ciafaloni:2001mu,Ciafaloni:2005fm} has still to be investigated. 

If on one hand the higher luminosity will improve the experimental accuracy, on the other hand 
the increase of the center of mass energy will allow to probe more and more extreme regions of 
the phase space, such as the ones characterized by a large jet multiplicity at high $p_T$ 
and missing transverse energy (also with additional hard isolated leptons) or by the production 
of boosted objects in association with jets, that are of great interest for both the direct new physics 
searches and also for quite challenging Standard Model measurements (see for instance the results 
of Refs.~\cite{Aad:2014qxa,Khachatryan:2014uva,Aad:2013ysa,Khachatryan:2014zya,Aad:2014rta,Khachatryan:2015ira} 
for $W+$~multi-jet and $Z+$~multi-jet production, as possible examples of direct new physics 
searches at the LHC we refer to~\cite{Chatrchyan:2014lfa,Aad:2013wta,CMS:2014exa,Aad:2015mia}). 
In these extreme kinematical 
configurations one loop electroweak corrections are dominated by double and single logarithms 
of the typical energy scale of the process over the gauge boson masses (the so called Sudakov 
logarithms~\cite{Sudakov:1954sw,Beccaria:1998qe}). 
The Sudakov logarithms basically correspond to the infrared regime of the virtual 
one loop weak corrections, where the gauge boson masses can be neglected if compared to the other 
energy scales involved. In the Sudakov limit the weak boson masses 
play the role of physical regulators for the \emph{infrared} 
singularities of the loop diagrams, which in turn  manifest themselves as logarithms of the 
energy scale over the \emph{infrared} cutoffs $M_W$ and $M_Z$ 
~\cite{Ciafaloni:1998xg,Beccaria:1999fk,Ciafaloni:2000df,Ciafaloni:2000rp,Ciafaloni:2001vt,Fadin:1999bq,Stirling:2012ak,
Denner:2000jv,Denner:2001gw}. 
The Sudakov corrections are negative and usually have a moderate impact on the total cross 
sections. However these corrections are sizable in particular in the tails of the $p_T$ 
or $H_T$ distributions~\cite{Chiesa:2013yma,Butterworth:2014efa} where they can be of the order 
of several tens of percent ($-40$\%, $-50$\%) at the 14 TeV LHC and 
become even larger at the center of mass energies of the proposed future hadron colliders 
such as the High-Energy LHC or the hh-FCC~\cite{Mishra:2013una,Campbell:2013qaa}. 
These large negative effects should be included 
in the theoretical predictions and are a clear indication of the role that the  
$\mathcal{O}( \alpha )$ corrections will play at high energy hadron colliders 
(also for processes with many final state particles).

\subsection{Electroweak corrections for the LHC: quest for automation}
\label{subsect:introautomation}
At present full one loop electroweak corrections are available only for a limited 
class of LHC processes, such as charged and neutral Drell-Yan, 
$V+1$~jet ($V=Z$, $W$, $\gamma$)~\cite{Maina:2004rb,Kuhn:2005gv,Kuhn:2005az,Kuhn:2007cv}, 
dilepton+jets~\cite{Denner:2009gj,Denner:2011vu,Denner:2012ts}, 
single top~\cite{Beccaria:2006ir,Beccaria:2008av,Bardin:2010mz}, 
$t \overline{t}$~\cite{Beenakker:1993yr,Kuhn:2005it,Moretti:2006nf,Bernreuther:2006vg,Kuhn:2006vh,Hollik:2007sw,Bernreuther:2008md}, 
dijet~\cite{Moretti:2005ut,Moretti:2006ea,Dittmaier:2012kx}, 
$Z/W+H$~\cite{Denner:2011id,Ciccolini:2003jy} (including the $Z/W$ decay products), $H$ 
production in vector boson fusion~\cite{Ciccolini:2007ec,Ciccolini:2007jr}, 
$V V'$ (with on-shell vector bosons or in pole approximation)~
\cite{Accomando:2005ra,Bierweiler:2012kw,Billoni:2013aba,Bierweiler:2013dja,Gieseke:2014gka,Baglio:2013toa},
$W W+1$~jet~\cite{Wei-Hua:2015gaa}, $WZZ$~\cite{Nhung:2013jta}, $WWZ$~\cite{Yong-Bai:2015xna},
$H \to 4f$~\cite{Bredenstein:2006ha,Boselli:2015aha}, $W \gamma$ production~\cite{Denner:2014bna}, 
$t \overline{t} +H$~\cite{Frixione:2014qaa,Yu:2014cka}. 
More recently,  exact $\mathcal{O}( \alpha )$ corrections to 
$Z(\to l \overline{l})+2$~jets~\cite{Denner:2014ina}, $W+n$~jets ($n \le 3$)~\cite{Kallweit:2014xda} and 
$t \overline{t} +H/Z/W$~\cite{Frixione:2014qaa,Frixione:2015zaa}.
Except for the last three examples, 
all the above computations have been carried out on a process-specific approach  
and are basically limited to processes with four or at most five 
external particles: the main reason of this limitation is the fact that the 
computational complexity grows considerably as the number of external leg 
particles increases.

As already stressed in the previous paragraphs, in order to fully exploit 
the physics potential of the second run of the LHC, the computation of the 
NLO electroweak corrections will be mandatory  for a very large number of 
processes (also with large final state multiplicities). It should be clear 
that in this context a process-specific approach for NLO EW calculations 
is no longer feasible and more general and process-independent techniques 
should be developed. 

As far as only the Sudakov regime is concerned, the universality of the infrared 
limit of weak corrections can be exploited to develop general algorithms for 
the calculation of the EW corrections in the logarithmic approximation
~\cite{Denner:2000jv,Denner:2001gw,Fadin:1999bq,Chiu:2008vv}. 
Following this approach, the Sudakov corrections to 
diboson~\cite{Accomando:2004de,Accomando:2005ra,Accomando:2005xp,Fuhrer:2010eu}, 
vector boson plus multi-jets~\cite{Chiesa:2013yma,Becher:2013zua}, 
$t \overline{t}$+jets~\cite{Manohar:2012rs}, $H$~\cite{Fuhrer:2010vi,Siringo:2012mi} 
and $H+$jet~\cite{Becher:2013zua} 
production have been computed pointing out further the phenomenological impact of 
the EW corrections at high energies. Order $\alpha$ corrections to dijet, Drell-Yan 
and $t\overline{t}$ production have been recently included in the {\sc{MCFM}} Monte 
Carlo program~\cite{Campbell:2010ff}: both the Sudakov approximation and the full 
one loop corrections have been implemented in order to provide a tool for the fast 
evaluation of the approximated $\mathcal{O}(\alpha)$ corrections that also allows 
to asses the validity of the approximated results~\cite{Campbell:2015vua}.
 
Although the Sudakov corrections are universal, this is not the case for the full 
NLO electroweak corrections that are supposed to work also for energy scales 
well below the Sudakov regime. Indeed, the non-logarithmic parts of the corrections are 
strictly process dependent. Despite the fact that $\mathcal{O}( \alpha )$ corrections 
are not universal, what could be cast in a process-independent form is the 
way these corrections are computed, namely the algorithms used to compute 
virtual loop diagrams, to perform the subtraction of IR singularities, to 
include ultraviolet renormalization and to compute real corrections. This is 
the first mandatory step in the direction of the automation of NLO EW 
calculations.

The development of process-independent algorithms to compute radiative corrections 
was the starting point also for the automation of QCD NLO corrections. Even if 
in principle the strategies developed for one loop QCD calculations can be extended 
to cover also the case of EW corrections, this generalization is not trivial 
because of several specific aspects of electroweak interactions, such 
as for example the chiral structure of weak interactions (which requires a 
consistent treatment of $\gamma^5$ matrices in loop diagrams), the treatment 
of unstable particles (that leads to the introduction of complex masses and -as 
result- of complex couplings), the highly non trivial interplay between QED and 
QCD-like singularities or the presence of very different mass scales in loop 
diagrams. On the real emission side the presence of several mass scales and unstable
particles typically leads to a phase space with a more involved structure containing
Breit-Wigner peaks compared for instance to massless QCD.

Quite recently the issue of automation of one loop electroweak corrections 
has been addressed by several collaborations and tools like {\sc{MadGraph5\textunderscore aMC@NLO}}, 
{\sc{OpenLoops}} and {\sc{RECOLA}} have been developed and applied to perform phenomenological 
studies on the impact of the NLO EW corrections to the processes 
$pp \to t \overline{t}+ \{ H, \, Z, \, W \}$, $pp \to W+n$~jets ($n=1,2,3$) and 
$pp \to Z(l \overline{l})+2$~jets, respectively. 

In Ref.~\cite{Kallweit:2014xda} the factorized NLO QCD+EW corrections to the 
processes $pp \to W^+ +$1, 2 and 3~jets have been computed using 
{\sc{OpenLoops}} interfaced with {\sc{MUNICH}}  
and {\sc{SHERPA}}. {\sc{OpenLoops}} computes LO and NLO amplitudes by means 
of four-dimensional recursion relations applied at the level of Feynman diagrams. 
In particular, one loop diagrams are written in terms of tensor integrals multiplied 
by coefficients that are determined numerically. The tensor integrals are then 
reduced to scalar integrals by means of the {\sc{COLLIER}} library~\cite{Denner:2014gla} which 
implements the algorithms of Refs.~\cite{Denner:2002ii,Denner:2005nn}. Rational terms and renormalization 
counterterms are included as additional tree-level Feynman rules. 
{\sc{MUNICH}} and {\sc{SHERPA}} compute real QCD/QED corrections and 
perform the subtraction of infrared singularities 
within the dipole subtraction scheme~\cite{Catani:1996vz,Catani:2002hc}. 
 
In Ref.~\cite{Denner:2014ina} the one loop corrections of absolute order $ \alpha_S^2 \alpha^3 $ 
to $Z+2$~jets have been computed using the code {\sc{RECOLA}}. 
Also in {\sc{RECOLA}} the computation 
of LO and NLO amplitudes is based on recursion relations which however  
act on off-shell currents (that correspond to sum of Feynman sub-
diagrams), leading to a different treatment of color with respect to 
{\sc{OpenLoops}}. As in {\sc{OpenLoops}}, the algorithm constructs numerically 
in four dimensions the coefficients of the tensor integrals that are then 
reduced using the {\sc{COLLIER}} library. Rational terms (as well as UV 
renormalization counterterms) are introduced at the level of LO 
amplitude generation, while IR singularities are canceled in the dipole 
scheme.  

One loop electroweak and QCD corrections to the processes $pp \to t \overline{t}+  H$, 
$pp \to t \overline{t}+  W$ and $pp \to t \overline{t}+  Z$ have been computed in the 
{\sc{MadGraph5\textunderscore aMC@NLO}} framework in Refs.~\cite{Frixione:2014qaa,Frixione:2015zaa}. 
IR singularities are subtracted in the FKS~\cite{Frixione:1995ms,Frixione:1997np} 
scheme implemented in {\sc{MADFKS}}~\cite{Frederix:2009yq}, while one loop amplitudes are 
generated by {\sc{MADLOOP}} which provides an independent implementation of the 
OpenLoops algorithm (where one loop integrals are computed either via tensor 
integral reduction or by means of the OPP method). 

As a final remark, it is worth noting that the automation 
of one loop EW corrections is still in its early stages, but rapidly evolving.

\section{Virtual corrections and scheme dependence}
\label{virtual}
In one formula, the basic ingredients of an NLO computation can be sketched as 
\begin{equation}
\sigma_n^{\rm NLO} = \sigma_n^{\rm Virt.} + \sigma_n^{\rm CT} + 
\sigma_n^{\rm Real, \; sub} + \int {\rm d} \Phi^{\rm rad} 
( \sigma_{n+1}^{\rm Real}-\sigma_{n+1}^{\rm sub} ) ,
\label{eq:nlo} 
\end{equation}
where $n$ is the number of external legs of the LO process under consideration. In Eq.~(\ref{eq:nlo}), 
$\sigma_n^{\rm Virt.}$ is the contribution of the diagrams with one virtual loop: these diagrams can be  
ultraviolet (UV) divergent when the number of propagators involved in the loop integral is smaller 
than four, so that a regularization procedure is needed in order to keep track of the UV singularities 
that will be removed by means of renormalization (in additive renormalization prescriptions, this is 
achieved by adding the appropriate counterterms -$\sigma_n^{\rm CT}$ in Eq.~(\ref{eq:nlo})- to the virtual 
corrections). Besides the UV singularities, in massless gauge theories such as QED or QCD, $\sigma_n^{\rm Virt.}$ 
can be also infrared (IR) divergent: the IR singularities of the virtual one loop corrections are 
cancelled by the corresponding IR singularities that appear in the computation of real NLO corrections. 
In Eq.~(\ref{eq:nlo}) we assume that the integral over the real phase space has been regularized by means 
of a subtraction procedure~\cite{Ellis:1980wv}, such as the Catani-Seymour~\cite{Catani:1996vz,Catani:2002hc} 
or the FKS subtraction~\cite{Frixione:1995ms,Frixione:1997np}, and 
$\sigma_n^{\rm Real, \; sub}$ represents the integrated contribution of the singular part of the real 
NLO corrections. In the context of NLO corrections to hadronic processes, additional singularities arise 
due to the collinear splitting of massless initial state partons: as these initial state singularities 
are universal, they can be absorbed in the parton distribution functions~(PDFs) and have to be 
consistently subtracted form the one loop calculation by adding to Eq.~(\ref{eq:nlo}) the corresponding 
collinear counterterms.

The commonly used regularization prescription for UV singularities is dimensional regularization 
(DR)~\cite{'tHooft:1972fi}. The basic idea of DR is to promote the number of space-time dimensions from $4$ 
to $D=4-2 \epsilon_{\rm UV} < 4$, so that the loop diagrams read 
\begin{equation}
\frac{(2 \pi \mu )^{4-D} }{i \pi^2}
\int {\rm d}^D q 
\frac{ \mathcal{N} (q,\{ p_i, \psi(p_i) \} )}{ \prod_{l=0}^N \left[ (q -r_l)^2 - m_l^2 \right] }, 
\label{eq:loopint}
\end{equation} 
where $q$ is the loop momentum, $\{ p_i, \psi(p_i) \}$ schematically denote the external momenta 
(considered as incoming) and their spinors or polarization vectors, $r_l=\sum_{i=1,l} p_i$, $m_l$ is 
the mass of the particle in $l$-th loop propagator and $\mu$ is a mass parameter introduced in order 
to keep the dimension of the integral fixed for varying $D$. In DR singularities appear as 
$1 / \epsilon_{\rm UV}$ poles in the limit $D \to 4$. In Eq.~(\ref{eq:loopint}) the full Lorentz and 
Dirac structure of the loop diagram is included in the numerator $\mathcal{N}$. In QCD calculations, DR 
is the standard regularization prescription also for IR singularities (even if $D=4-2 \epsilon_{\rm IR} > 4$ 
and $\epsilon_{\rm IR} = - \epsilon_{\rm UV}$). In the context of NLO electroweak computations, the 
traditional IR regularization scheme is mass regularization (\emph{i.e.} IR singularities are regularized 
by giving an unphysical mass to the photon and the massless fermions), however DR is becoming the new 
standard procedure in order to treat the IR limit of QED and QCD on the same footing (this is 
of particular importance not only for the computation of mixed QCD and electroweak corrections, but also 
for the extension to electroweak computations of NLO+parton shower matching algorithms originally 
developed for QCD calculations~\cite{Nason:2004rx,Frixione:2002ik,Barze:2012tt,Barze':2013yca,Barze:2014zba}). 

In the original formulation of DR in Ref.~\cite{'tHooft:1972fi}, not only the loop momentum but also all the other 
momenta become $D$-dimensional together with the Lorentz covariants (such as $g^{\mu \nu}$ and $\gamma^{\mu}$), 
which become formal objects obeying specific algebraic relations. This kind of continuation to $D$ dimension 
is consistent and well defined for non-chiral theories, while problems arise in the treatment of the $\gamma^5$ 
matrix and the $\varepsilon_{\mu \nu \rho \sigma}$ antisymmetric tensor. More precisely, already in 
Ref.~\cite{'tHooft:1972fi} has been pointed out that the main properties of the $\gamma^5$ matrix 
in four dimensions (i.e. $\{ \gamma^5, \gamma^{\mu} \}=0$, 
${\rm Tr}(\gamma^5 \gamma^{\mu} \gamma^{\nu} \gamma^{\rho} \gamma^{\sigma})= 4i \varepsilon_{\mu\nu\rho\sigma}$ 
together with the trace cyclicity), in $D$ dimension could lead to algebraic inconsistencies in presence of 
anomalies. In the literature, several different recipes to handle the $\gamma^5$ matrix in the context of DR 
have been proposed:
\begin{itemize}
\item{\emph{'t Hooft $\gamma^5$ prescription.} This $\gamma^5$ prescription has been proposed by 
't Hooft and Veltman~\cite{'tHooft:1972fi}, Akyeampong and Delbourgo~\cite{Akyeampong:1973xi,Akyeampong:1973vk}  
and systematized by Breitenlohner and Maison~\cite{Breitenlohner:1977hr}. 
According to Ref.~\cite{'tHooft:1972fi}:
\begin{equation}
\gamma^5=i \gamma^0 \gamma^1 \gamma^2 \gamma^3 = \frac{i}{4!} \varepsilon_{\mu \nu \rho \sigma } 
\hat{ \gamma }^{ \mu } \hat{ \gamma }^{ \nu } \hat{ \gamma }^{ \rho } \hat{ \gamma }^{ \sigma } ,
\label{eq:g5thooft1}
\end{equation}
\begin{equation}
\{ \gamma^5,  \gamma^{\mu} \} \neq 0, \qquad  
\{ \gamma^5,  \hat{\gamma}^{\mu} \} = 0, \qquad 
\left[ \gamma^5,  \tilde{\gamma}^{\mu} \right] =0.
\label{eq:g5thooft2}
\end{equation}
In equation~(\ref{eq:g5thooft2}) we introduced the notation $\gamma^{\mu}=\hat{\gamma}^{\mu}+\tilde{\gamma}^{\mu}$, 
where $\hat{\gamma}^{\mu}$ and $\tilde{\gamma}^{\mu}$ represent $4$ and $D-4$ dimensional objects, respectively. 
Even if algebraically consistent, the 't Hooft $\gamma^5$ prescription leads to spurious anomalies that should 
be removed by introducing appropriate finite counterterms.
}
\item{\emph{KKS prescription}. The scheme proposed in 
Refs.~\cite{Korner:1989is,Kreimer:1993bh,Korner:1991sx} preserves the anticommutation 
relationship $\{ \gamma^5, \gamma^{\mu} \}=0$ and prevented cyclic property in Dirac traces to avoid algebraic 
inconsistency. A projection on four-dimensional subspaces is needed in the redefinition of trace operation: 
\begin{equation}
{\rm Tr} (\gamma^5 \gamma^{\mu} \gamma^{\nu} \gamma^{\rho} \gamma^{\sigma})= 
{\rm Tr} (\mathcal{P} ( \gamma^5) \gamma^{\mu} \gamma^{\nu} \gamma^{\rho} \gamma^{\sigma}), \qquad 
\mathcal{P} ( \gamma^5)= \frac{i}{4!} \varepsilon_{\mu \nu \rho \sigma } 
\hat{ \gamma }^{ \mu } \hat{ \gamma }^{ \nu } \hat{ \gamma }^{ \rho } \hat{ \gamma }^{ \sigma }.
\label{eq:kks}
\end{equation}
Since no cyclic permutation of Dirac matrix inside traces is allowed, all the traces must be read 
starting from the same vertex (reading point prescription).
}
\item{\emph{Naive DR.} In this prescription $\{ \gamma^5, \gamma^{\mu} \}=0$, 
${\rm Tr}(\gamma^5 \gamma^{\mu} \gamma^{\nu} \gamma^{\rho} \gamma^{\sigma})= 4i \varepsilon_{\mu\nu\rho\sigma}$ 
and cyclic permutation of gamma matrices inside traces are allowed (see for 
instance Ref.~\cite{Chanowitz:1979zu}). Even though these properties can be incompatible, 
the resulting ambiguities are expected to cancel in non-anomalous 
theories~\cite{Hahn:1998yk,Schubert:1988ke,Ferrari:1994ct}. We will take the formulae 
for the electroweak counterterms from Ref.~\cite{Denner:1991kt}, where the naive prescription has been used: accordingly 
we will use this prescription in the examples of subsection~\ref{susect:examples} for our CDR (conventional DR) results.}
\end{itemize}
These three different prescriptions for the treatment of $\gamma^5$ are three 
possible examples of different realizations of DR: 
all of them share the same definition of the denominator of the loop diagram in Eq.~(\ref{eq:loopint}), while they 
differ in the treatment of the numerator of Eq.~(\ref{eq:loopint}). Different choices about the numerator 
algebra in Eq.~(\ref{eq:loopint}) concerning the dimension of the momenta, the Dirac matrices or the 
spinors and polarization vectors, define different DR schemes, such as for example the conventional 
DR scheme~(CDR)~\cite{'tHooft:1972fi,Bollini:1972ui,Ashmore:1972uj,Cicuta:1972jf} (or some variant thereof), 
the 't Hooft Veltman (HV) scheme~\cite{'tHooft:1972fi} or the Four Dimensional Helicity scheme (FDH)~\cite{Bern:1992em,Kunszt:1993sd}. 
Dimensional Reduction (DRED)~\cite{Siegel:1979wq,Siegel:1980qs,Jack:1994bn,Stockinger:2005gx} can 
be regarded as a DR scheme in which all the objects appearing in the numerator of Eq.~(\ref{eq:loopint}) are 
four dimensional (except, of course, the loop momentum): this is the regularization scheme used by {\sc{GoSam}}
for the 
computation of one loop diagrams and it is particularly appealing because it allows to work out the numerator 
of Eq.~(\ref{eq:loopint}) using four dimensional relations plus some trivial algebra to treat the loop momentum 
components in the extra dimensions without introducing special Feynman rules (for the finite integrals
involved see for example Ref.~\cite{delAguila:2004nf}). 

Different DR schemes lead to different results for the computation of one loop diagrams. The differences consist 
of $\epsilon$-dependent terms, that vanish in the $D \to 4$ limit unless they multiply the $1 / \epsilon $ poles, 
giving rise to finite terms (usually called rational terms in the literature). The main purpose of the present 
paper is to show how the automation of NLO electroweak corrections can be achieved starting from 
existing tools, such as {\sc{GoSam}} (for the computation of one loop matrix elements) and 
{\sc{MadDipole}} (for the IR subtraction). If we recall  
Eq.~(\ref{eq:nlo}), one of the technical issues we have to face is the matching between the $\sigma_n^{\rm Virt.}$ 
term computed by {\sc{GoSam}} in DRED and the contributions coming from renormalization and from the integrated 
dipoles ($\sigma_n^{\rm CT}$ and $\sigma_n^{\rm Real, \; sub}$, respectively) that are usually computed in the 
CDR scheme in the literature. 
Although the different building blocks of an NLO calculation are  
regularization scheme dependent, as pointed out in Ref.~\cite{Catani:1996pk} 
unitarity implies that the regularization scheme dependence vanishes 
in the sum of $\sigma_n^{\rm Virt.}$, $\sigma_n^{\rm CT}$ and 
$\sigma_n^{\rm Real, \; sub}$, once these three terms have been computed 
within the same regularization scheme. 
For the counterterms part, in particular, we implemented the formulae of 
Ref.~\cite{Denner:1991kt} for the on-shell renormalization scheme and derived the conversion factors from CDR to DRED. 
The infrared subtraction procedure implemented in {\sc{MadDipole}} is based on 
the Catani-Seymour dipole formalism: the CDR to DRED conversion factors 
for the integrated dipoles have been taken from Ref.~\cite{Catani:1996pk} 
(with a straightforward generalization to the case of QED radiation). 
In subsection~\ref{subsect:virtrenorm} we give a short overview about the electroweak renormalization 
at one loop in the on-shell scheme, we also explain our treatment of massless fermions and list the relevant 
regularization scheme conversion factors for the electroweak counterterms. 
The subtraction of IR singularities is described 
in subsection~\ref{susec:elinop}, while in subsection \ref{scheme} we address the issue of the regularization 
scheme dependence of the infrared subtraction procedure. As examples on how the things work, 
in subsection~\ref{susect:examples} 
we show the CDR and DRED results for the NLO electroweak corrections to three simple processes:  
$Z\to \nu \bar{\nu}$, $W^+ \to u \bar{d}$ and $u \bar{u} \to c \bar{c}$. 
In particular, with the examples we show the regularization scheme 
independence of the sum of virtual one loop corrections, renormalization 
counterterms and integrated dipoles when these three contributions are 
consistently computed in the CDR or DRED scheme. 

\subsection{Electroweak renormalization and counterterms}
\label{subsect:virtrenorm}
We use the on-shell scheme~\cite{Ross:1973fp,Sirlin:1980nh,Marciano:1980pb,Sirlin:1981yz,
Sakakibara:1980hw,Fleischer:1980ub,Bardin:1980fe,Bardin:1981sv,Aoki:1982ed,Bohm:1986rj} for the electroweak 
renormalization at one loop. For the sake of simplicity, we employ the notation of Ref.~\cite{Denner:1991kt} 
and refer to Appendix~B of~\cite{Denner:1991kt} for the explicit expressions of the self energies involved 
in the formulas for the electroweak counterterms. 
Each free parameter of the Standard Model Lagrangian is considered as a bare parameter and it is split 
into its renormalized version and the corresponding counterterm, namely:
\begin{eqnarray}
e_0 & = &e +\delta e =(1+ \delta Z_e)e , \qquad m_{f,0}  =  m_f + \delta m_f \nonumber \\
M^2_{W,0} & = & M^2_W + \delta M^2_W, \qquad M^2_{Z,0}  =  M^2_Z + \delta M^2_Z,
 \qquad M^2_{H,0}  =  M^2_H + \delta M^2_H ,
\label{eq:defdeltac}
\end{eqnarray} 
(where $f$ stands for a generic fermion field), then for each external field we introduce the 
corresponding wave function renormalization counterterm:
\begin{eqnarray}
W_0 & = & ( 1+\frac{1}{2}\delta Z_W)W, \qquad H_0  =  ( 1+\frac{1}{2}\delta Z_H)H, \nonumber \\
f_0^L & = & ( 1+\frac{1}{2}\delta Z_f^L) f^L, \qquad f_0^R  = ( 1+\frac{1}{2}\delta Z_f^R) f^R, \nonumber \\
Z_0 & = & ( 1+\frac{1}{2}\delta Z_{ZZ}) Z +\frac{1}{2}\delta Z_{ZA} A, \nonumber \\
A_0 & = & ( 1+\frac{1}{2}\delta Z_{AA}) A +\frac{1}{2}\delta Z_{AZ} Z.
\label{eq:defdeltaf}
\end{eqnarray} 
This way the Lagrangian can be split into the basic Lagrangian and the counterterm Lagrangian 
$\mathcal{L}_0 = \mathcal{L} + \delta \mathcal{L}$, where $\delta \mathcal{L}$ provides the Feynman rules for 
the counterterms.

In the on-shell scheme the form of the renormalization counterterms is fixed by the following 
requirements:
\begin{itemize}
\item{the renormalized masses are the real parts of the poles of the propagators at one loop;}
\item{the residues of the renormalized propagators are equal to one;}
\item{the renormalized electric charge is obtained from the $ee \gamma$ vertex in the Thomson limit;}
\item{the renormalized Higgs tadpole is set to zero.}
\end{itemize}
With the above conditions the one loop EW counterterms for fermions, gauge bosons and EW couplings read:

\begin{alignat}{1}
\label{eq:rencond2}
\delta m_{f} & 
=\frac{m_{f}}{2}\tilde{\textrm{Re}}\left(\Sigma_{ii}^{f,L}\left(m_{f}^{2}\right)
+\Sigma_{ii}^{f,R}\left(m_{f}^{2}\right)+2\Sigma_{ii}^{f,S}\left(m_{f}^{2}\right)\right), \nonumber\\
\delta Z_{ii}^{f,L} & 
=-\tilde{\textrm{Re}}\Sigma_{ii}^{f,L}  \left(m_{f}^{2}\right)
 -m_{f}^{2}\frac{\partial}{\partial p^{2}}\left(\tilde{\textrm{Re}} \Sigma_{ii}^{f,L}\left(p^{2}\right)
+\tilde{\textrm{Re}} \Sigma_{ii}^{f,R}\left(p^{2}\right)
+2\tilde{\textrm{Re}} \Sigma_{ii}^{f,S}\left(p^{2}\right)\right)\Bigg|_{p^{2}=m_{f}^{2}}, \nonumber\\
\delta Z_{ii}^{f,R} & 
=-\tilde{\textrm{Re}}\Sigma_{ii}^{f,R}\left(m_{f}^{2}\right)
-m_{f}^{2}\frac{\partial}{\partial p^{2}}\left(\tilde{\textrm{Re}} \Sigma_{ii}^{f,L}\left(p^{2}\right)
+\tilde{\textrm{Re}} \Sigma_{ii}^{f,R}\left(p^{2}\right)
+2\tilde{\textrm{Re}}\Sigma_{ii}^{f,S}\left(p^{2}\right)\right)\Bigg|_{p^{2}=m_{f}^{2}},
\end{alignat}

\begin{alignat}{2}
\label{eq:rencond1}
\delta M_{W}^{2} & =\tilde{\textrm{Re}}\Sigma_{T}^{W}\left(M_{W}^{2}\right), & \delta Z_{W} & = -\frac{\partial}{\partial k^{2}}\tilde{\textrm{Re}}\Sigma_{T}^{W}\left(k{}^{2}\right)\Bigg|_{k^{2}=M_{W}^{2}}, \nonumber \\
\delta M_{Z}^{2} & =\tilde{\textrm{Re}}\Sigma_{T}^{ZZ}\left(M_{Z}^{2}\right), & \delta Z_{ZZ} & =-\frac{\partial}{\partial k^{2}}\tilde{\textrm{Re}}\Sigma_{T}^{ZZ}\left(k^{2}\right)\Bigg|_{k^{2}=M_{Z}^{2}}, \nonumber \\
\delta Z_{AA} & =-\frac{\partial}{\partial k^{2}}\Sigma_{T}^{AA}\left(k^{2}\right)\Bigg|_{k^{2}=0}, \nonumber \\
\delta Z_{AZ} & =-2\tilde{\textrm{Re}}\frac{\Sigma_{T}^{AZ}\left(M_{Z}^{2}\right)}{M_{Z}^{2}}, & \delta Z_{ZA} & =2\frac{\Sigma_{T}^{AZ}\left(0\right)}{M_{Z}^{2}}, \nonumber \\
\delta M_{H}^{2} & =\tilde{\textrm{Re}}\Sigma^{H}\left(M_{H}^{2}\right), & \delta Z_{H} & =-\frac{\partial}{\partial k^{2}}\tilde{\textrm{Re}}\Sigma^{H}\left(k^{2}\right)\Bigg|_{k^{2}=M_{H}^{2}},
\end{alignat}

\begin{alignat}{1}
\label{eq:rencond3}
\frac{\delta e}{e} & =-\frac{1}{2}\delta Z_{AA}-\frac{s_{W}}{c_{W}}\frac{1}{2}\delta Z_{ZA}
=\frac{1}{2}\frac{\partial}{\partial k^{2}}\Sigma_{T}^{AA}\left(k^{2}\right)\Bigg|_{k^{2}=0}-
\frac{s_{W}}{c_{W}}\frac{\Sigma_{T}^{AZ}\left(0\right)}{M_{Z}^{2}},\nonumber\\
\frac{\delta c_{W}}{c_{W}} & 
=\frac{1}{2}\left(\frac{\delta M_{W}^{2}}{M_{W}^{2}}-\frac{\delta M_{Z}^{2}}{M_{Z}^{2}}\right)
=\frac{1}{2}\tilde{\textrm{Re}}\left(\frac{\Sigma_{T}^{W}\left(M_{W}^{2}\right)}{M_{W}^{2}}
-\frac{\Sigma_{T}^{ZZ}\left(M_{Z}^{2}\right)}{M_{Z}^{2}}\right),\nonumber\\
\frac{\delta s_{W}}{s_{W}} & 
=-\frac{c_{W}^{2}}{s_{W}^{2}}\frac{\delta c_{W}}{c_{W}}=
-\frac{1}{2}\frac{c_{W}^{2}}{s_{W}^{2}}\tilde{\textrm{Re}}\left(\frac{\Sigma_{T}^{W}\left(M_{W}^{2}\right)}{M_{W}^{2}}
-\frac{\Sigma_{T}^{ZZ}\left(M_{Z}^{2}\right)}{M_{Z}^{2}}\right),\nonumber\\
\delta t & = -T^H.
\end{alignat}
Where the notation $\tilde{\textrm{Re}}$ means that the real part is taken only for the scalar functions contained in the 
unrenormalized self energies. As in our computation the CKM matrix $V_{\rm CKM }$ is chosen to be diagonal and no renormalization 
of $V_{\rm CKM }$ is needed, we do not discuss here the renormalization conditions for the quark mixing matrix.

In the following we list the conversion rules from the CDR to the DRED regularization schemes. 
For electroweak couplings and gauge bosons the conversion rules read:   
\begin{eqnarray}
\label{eq:convBOS}
\delta Z_e^{\rm DRED}    & = & \delta Z_e^{\rm CDR}    + \frac{\alpha}{4\pi} \frac{1}{3}, \qquad
\delta Z_{ZA}^{\rm DRED}  = \delta Z_{ZA}^{\rm CDR}, \qquad \qquad \qquad
\delta Z_{AZ}^{\rm DRED}  = \delta Z_{AZ}^{\rm CDR} + \frac{\alpha}{4\pi} \frac{4}{3}\frac{c_W}{s_W} \nonumber \\
\delta Z_{AA}^{\rm DRED} & = & \delta Z_{AA}^{\rm CDR} - \frac{\alpha}{4\pi} \frac{2}{3}, \qquad
\delta Z_{ZZ}^{\rm DRED}  = \delta Z_{ZZ}^{\rm CDR} - \frac{\alpha}{4\pi} \frac{2}{3}\frac{c_W^2}{s_W^2}, \qquad
\delta Z_{W }^{\rm DRED}  = \delta Z_{W }^{\rm CDR} - \frac{\alpha}{4\pi} \frac{2}{3 s_W^2}, \nonumber \\
\delta M_{W}^{2 \, \rm DRED} & = & \delta M_{W}^{2 \, \rm CDR} + \frac{\alpha}{4\pi} \frac{2}{3} \frac{M_W^2}{s_W^2}, \qquad
\delta M_{Z}^{2 \, \rm DRED}   =   \delta M_{Z}^{2 \, \rm CDR} + \frac{\alpha}{4\pi} \frac{2}{3} \frac{M_Z^2 c_W^2} {s_W^2}, \nonumber \\
\delta Z_{H }^{\rm DRED}  &=& \delta Z_{H }^{\rm CDR}, \qquad 
\delta M_{H}^{2 \, \rm DRED}   =   \delta M_{H}^{2 \, \rm CDR} + \frac{\alpha}{4\pi} \frac{3(2c_W^2 M_W^2 +M_Z^2)}{2 c_W^2 s_W^2}.
\end{eqnarray}
The CDR to DRED conversion rules for up-type quarks, down-type quarks and leptons, respectively, are:
\begin{eqnarray}
\label{eq:convUquark}
\delta m_u^{  \, \rm DRED} &=& \delta m_u^{  \, \rm CDR} +\frac{\alpha}{4\pi} m_u \frac{(3-4s_W^2)^2+2c_W^2(9+8s_W^2)}{72 c_W^2 s_W^2}, \nonumber \\
\delta Z_u^{L \, \rm DRED} &=& \delta Z_u^{L \, \rm CDR} -\frac{\alpha}{4\pi}    \frac{(3-4s_W^2)^2+2c_W^2(9+8s_W^2)}{36 c_W^2 s_W^2}, \nonumber \\
\delta Z_u^{R \, \rm DRED} &=& \delta Z_u^{R \, \rm CDR} -\frac{\alpha}{4\pi}    \frac{4}{9 c_W^2};
\end{eqnarray}
\begin{eqnarray}
\label{eq:convDquark}
\delta m_d^{  \, \rm DRED} &=& \delta m_d^{  \, \rm CDR} +\frac{\alpha}{4\pi}m_d \frac{9+4s_W^4+2c_W^2(9+2s_W^2)}{72 c_W^2 s_W^2}, \nonumber \\
\delta Z_d^{L \, \rm DRED} &=& \delta Z_d^{L \, \rm CDR} -\frac{\alpha}{4\pi}    \frac{(3-2s_W^2)^2+2c_W^2(9+2s_W^2)}{36 c_W^2 s_W^2}, \nonumber \\
\delta Z_d^{R \, \rm DRED} &=& \delta Z_d^{R \, \rm CDR} -\frac{\alpha}{4\pi}    \frac{1}{9 c_W^2};
\end{eqnarray}
\begin{eqnarray}
\label{eq:convlept}
\delta m_e^{  \, \rm DRED} &=& \delta m_e^{  \, \rm CDR} +\frac{\alpha}{4\pi}m_e \frac{1+4s_W^2(s_W^2-2)+c_W^2(2+4s_W^2)}{8 c_W^2 s_W^2}, \nonumber \\
\delta Z_e^{L \, \rm DRED} &=& \delta Z_e^{L \, \rm CDR} -\frac{\alpha}{4\pi}    \frac{1+2c_W^2}{4 c_W^2 s_W^2}, \nonumber \\
\delta Z_e^{R \, \rm DRED} &=& \delta Z_e^{R \, \rm CDR} -\frac{\alpha}{4\pi}    \frac{1}{ c_W^2}, \nonumber \\
\delta Z_{\nu}^{L \, \rm DRED} &=& \delta Z_{\nu}^{L \, \rm CDR} -\frac{\alpha}{4\pi}    \frac{1+2c_W^2}{4 c_W^2 s_W^2}.
\end{eqnarray}

As in our numerical examples we deal with external massless quarks, following for example Ref.~\cite{Barze:2012tt},  
we adopt a slightly modified version of the on-shell scheme where the light quark masses are set to zero in the 
corresponding external wave function renormalization counterterms (and mass counterterms for light quarks vanish). 
In this approach eqs.~(\ref{eq:convUquark})-(\ref{eq:convDquark}) 
for massless up and down-type quarks in particular become:
\begin{eqnarray}
\label{eq:convUquark2}
\delta Z_u^{L \, \rm DRED} &=& \delta Z_u^{L \, \rm CDR} -\frac{\alpha}{4\pi}    \frac{(3-4s_W^2)^2+18 c_W^2}{36 c_W^2 s_W^2}, \nonumber \\
\delta Z_u^{R \, \rm DRED} &=& \delta Z_u^{R \, \rm CDR} -\frac{\alpha}{4\pi}    \frac{4 s_W^2}{9 c_W^2};
\end{eqnarray}
\begin{eqnarray}
\label{eq:convDquark2}
\delta Z_d^{L \, \rm DRED} &=& \delta Z_d^{L \, \rm CDR} -\frac{\alpha}{4\pi}    \frac{(3-2s_W^2)^2+18 c_W^2}{36 c_W^2 s_W^2}, \nonumber \\
\delta Z_d^{R \, \rm DRED} &=& \delta Z_d^{R \, \rm CDR} -\frac{\alpha}{4\pi}    \frac{ s_W^2}{9 c_W^2}.
\end{eqnarray}
Large logarithms of the fermion masses appear in the counterterms $\delta Z_e$ and $\delta Z_{AA}$. They are basically 
related to the running of $\alpha$ from the Thomson limit to the electroweak scale. These terms are singular in the 
limit $m_q \to 0$, however they can be systematically cancelled by following the recommendation of the \emph{EW dictionary} 
of Ref.~\cite{Butterworth:2014efa}, namely by choosing $\alpha(0)$ as coupling constant for the LO vertices involving 
external photons and $\alpha(M_Z^2)$ or $\alpha_{G_{\mu}}$ for the remaining LO vertices. In the former case the 
logarithms of the light fermion masses cancel in the combination $\delta Z_e -\frac{1}{2} \delta Z_{AA}$, while 
in the latter they vanish as a result of the finite shifts  
$\delta Z^e |_{ \alpha( M_Z^2 ) } \to \delta Z^e |_{ \alpha(0)} - \frac{1}{2} \Delta \alpha(M_Z^2) $,
$\delta Z^e |_{G_{ \mu }} \to \delta Z^e |_{ \alpha(0)} - \frac{1}{2} \Delta r$ 
\cite{Denner:1991kt,Dittmaier:2001ay,Dittmaier:2012kx}. The remaining dependence on the light quark masses in the 
counterterms vanishes in the $m_q \to 0$ limit. Indeed, in this approach light 
quark masses can be safely neglected from the very beginning.  

We conclude this section with a short remark on unstable particles. 
In general the description of resonances in perturbation theory requires 
a Dyson summation of the self energy insertions. This leads to a mixing 
of perturbative orders that could break gauge invariance. In the context 
on one loop electroweak corrections unstable particles are usually treated 
within the so-called complex mass scheme (CMS)~\cite{Denner:1999gp,Denner:2005fg,Denner:2006ic}. 
In the CMS unstable particles masses are promoted to complex numbers 
through the replacement
\begin{equation}
 m_{V}^2 \to \mu_{V}^2 = m_{V}^2 -i m_{V} \Gamma_{V}\;,
\end{equation}
(where $\Gamma_{V}$ is the decay width of the unstable particle $V$) 
and the complex Weinberg angle is then defined via
\begin{equation}
 \cos^2 \theta_W = \frac{\mu_W^2}{\mu_Z^2}\;.
\end{equation}
Within the CMS renormalization can be performed in a modified version of the 
on shell scheme. The renormalized masses of the unstable particles are 
defined as the poles of the corresponding propagators in the complex plane. 
The resulting expressions for the counterterms are basically the same as in 
eqs.~(\ref{eq:rencond2}-\ref{eq:rencond3}) with the important difference 
that all the self energies involved become complex (via their dependence on 
the complex couplings and masses) and no real part is taken. In the CMS also 
the momentum flowing in the self energies is a complex variable. However, 
it is possible to realize a minimal version of the on shell renormalization 
in the CMS by expanding the self energies about the real arguments in such 
a way that one loop accuracy is retained. This in particular requires the 
introduction of an additional renormalization factor for the $W$ (and also 
the top) self energy in order to deal with the branch cut at $k^2=\mu^2_{W \,(t)}$ 
arising from the photonic loop diagrams. The resulting counterterms are still 
complex, but depend on real momenta.

\subsection{Electroweak insertion operator}
\label{susec:elinop}
In equation~(\ref{eq:nlo}) we briefly sketched the basic building blocks of an NLO calculation. 
We pointed out that virtual corrections $\sigma_n^{\rm Virt.}$ and real contributions $\sigma_{n+1}^{\rm Real}$ 
are separately infrared divergent and the IR singularities only cancel in the sum of the two 
terms (as far as infrared safe observables are considered and the initial state collinear singularities 
are properly absorbed in the PDFs in the case of hadronic processes). In Eq.~(\ref{eq:nlo}) 
we also introduced the infrared subtraction term $\sigma_{n+1}^{\rm sub}$ and its integrated 
counterpart $\sigma_{n}^{\rm Real, \; sub}$. $\sigma_{n+1}^{\rm sub}$ is a function of the $n+1$ kinematics 
that behaves as the matrix element for the real corrections in the soft/collinear limit: this 
way $\sigma_{n+1}^{\rm Real}-\sigma_{n+1}^{\rm sub}$ in Eq.~(\ref{eq:nlo}) is IR finite 
and can be integrated numerically. $\sigma_{n+1}^{\rm sub}$ should also have a simple 
expression, so that the IR subtraction term can be integrated analytically over the real 
radiation phase space: the resulting integral $\sigma_{n}^{\rm Real, \; sub}$ is a function 
of the tree level $n$~particle kinematics that has to be added back to the virtual corrections 
in order to cancel the IR poles of the one loop contributions.

Owing to the universality of the infrared limit of massless gauge theories like QCD it is 
possible to develop process independent algorithms to build the IR subtraction terms and 
of course their integrated counterparts (the dipole subtraction or the FKS methods are two 
examples of such algorithms).  
The infrared properties of one loop amplitudes are well known and in the context of QCD 
are usually described by a so called insertion operator~\cite{Catani:2000ef}, that 
corresponds to the integrated subtraction terms described above. 
These considerations can be easily translated to electroweak one loop amplitudes and the infrared
properties of such amplitudes are then described by an electroweak insertion operator.
The electroweak infrared
insertion operator is particularly useful as it allows to check the cancellation of double and single pole of the 
renormalized amplitude for a given phase space point. It can be obtained for instance from the results of
\cite{Catani:1996vz,Catani:2002hc} by, simply speaking, replacing the color factors by electric charges.
More specifically we start by defining the singular pieces as: 
\begin{equation}
 {\cal V}_{sing, QED}(s_{ik},m_i,m_k)=\frac{\alpha}{2\pi}\left(\frac{1}{\epsilon^2}{\cal V}_{\epsilon^2}(s_{ik},m_i,m_k)
+\frac{1}{\epsilon}{\cal V}_{\epsilon}(s_{ik},m_i,m_k)\right)\cdot |{\cal M}_{Born}|^2\;.
\end{equation}
The index $i$ denotes the combination of emitter and unresolved particle, the index $k$ denotes the spectator. The coefficients of the double and the single
pole depend on the masses and are given by
\begin{flalign}
 m_i>0,\; m_k>0:&\nonumber \\
& {\cal V}_{\epsilon^2}(s_{ik},m_i,m_k)=0 \nonumber \\
& {\cal V}_{\epsilon}(s_{ik},m_i,m_k)=\frac{\log\rho}{v_{ik}},\quad v_{ik}=\frac{\sqrt{\lambda(s_{ik},m^2_i,m^2_k)}}{s_{ik}-m^2_i-m^2_k},\;
\rho=\sqrt{\frac{1-v_{ik}}{1+v_ik}},\nonumber\\
& \qquad \qquad \quad \qquad \quad \lambda (x,y,z)=x^2+y^2+z^2-2xy-2yz-2xz,\nonumber\\
m_i>0,\; m_k=0:&\nonumber \\
&{\cal V}_{\epsilon^2}(s_{ik},m_i,0)=\frac{1}{2} \nonumber \\
&{\cal V}_{\epsilon}(s_{ik},m_i,0)= \frac{1}{2}\left(\frac{5}{2}+\log\frac{m^2_i}{s_{ik}}\right)\nonumber \\
m_i=0,\; m_k>0:&\nonumber \\
&{\cal V}_{\epsilon^2}(s_{ik},0,m_k)=\frac{1}{2} \nonumber \\
&{\cal V}_{\epsilon}(s_{ik},0,m_k)= \frac{1}{2}\left(\frac{5}{2}+\log\frac{m^2_k}{s_{ik}}\right)\nonumber \\
m_i=0,\; m_k=0:&\nonumber \\
&{\cal V}_{\epsilon^2}(s_{ik},0,0)=1 \nonumber \\
&{\cal V}_{\epsilon}(s_{ik},0,0)=\frac{3}{2} \;.
\end{flalign}
The insertion operator depends on the fact whether emitter/spectator are in the initial or final state. In particular, 
we have:\\
\begin{flalign}
 \text{initial-initial:}&\nonumber\\
&\mathbf{I}_{\epsilon^2}(s_{ik},m_i,m_k)=\frac{n_c\sigma_i q_i\sigma_kq_k}{\epsilon^2}{\cal V}_{\epsilon^2}(s_{ik},m_i,m_k)\cdot |{\cal M}_{Born}|^2\nonumber\\
&\mathbf{I}_{\epsilon}(s_{ik},m_i,m_k,\mu^2)=\frac{n_c\sigma_iq_i\sigma_kq_k}{\epsilon}{\cal V}_{\epsilon}(s_{ik},m_i,m_k)\left(\frac{3}{2} 
+ \log\frac{\mu^2}{s_{ik}}\right)\cdot |{\cal M}_{Born}|^2\nonumber\\
 \text{initial-final}\;/&\; \text{final-initial}\;/\; \text{final-final}:\nonumber\\
&\mathbf{I}_{\epsilon^2}(s_{ik},m_i,m_k)=\frac{n_c\sigma_iq_i\sigma_kq_k}{\epsilon^2}{\cal V}_{\epsilon^2}(s_{ik},m_i,m_k)\cdot |{\cal M}_{Born}|^2\nonumber\\
&\mathbf{I}_{\epsilon}(s_{ik},m_i,m_k,\mu^2)=\frac{n_c\sigma_iq_i\sigma_kq_k}{\epsilon}\left({\cal V}_{\epsilon}(s_{ik},m_i,m_k) 
+ \log\frac{\mu^2}{s_{ik}}{\cal V}_{\epsilon^2}(s_{ik},m_i,m_k) \right)\cdot |{\cal M}_{Born}|^2.
\label{eq:I}
\end{flalign}
The factors $q_i$, $q_k$ denote the fractional electric charges of emitter and spectator and $n_c$ is the number of 
colors that can occur in the 
splitting (i.e. either one or three). In addition, we follow Ref.\cite{Dittmaier:1999mb}
and introduce the sign factors $\sigma_{i,k}$ which are defined to be $+1$ for incoming fermions and outgoing anti-fermions 
and $-1$ for incoming anti-fermions and outgoing fermions. This can be extended to the case of a final state $W$-boson 
and yields a $+1$ for an external $W^{-}$ and a $-1$ for an external $W^{+}$. 
The sign factors ensure the conservation of the electric charge which is given by
\begin{equation}
\label{eq:signfactor}
 \sum_n \sigma_n\cdot q_n = 0\;.
\end{equation}

\subsection{Subtraction terms and scheme dependence}
\label{scheme}
Numerical simulations require the use of subtraction terms to render the real emission contribution finite.
One of the widely used subtraction methods in the context of electroweak calculation is the dipole formalism
\cite{Catani:1996vz} which has been adapted to electroweak calculations \cite{Dittmaier:1999mb,Dittmaier:2008md}.
In practice the subtraction methods usually also takes care of issues regarding the regularization scheme 
by adding appropriate additional terms to cancel the scheme dependence in the virtual corrections. 

As mentioned above, different regularization schemes lead to differences in the finite contribution for the virtual corrections. 
At the one-loop level the transition from one scheme to the other can be obtained by a simple shift that is proportional
to the Born contribution~\cite{Catani:1996pk}. Following the derivation of Ref.~\cite{Catani:1996pk}, in the case of photon emission 
one obtains a conversion term of the form: 
\begin{equation}
\label{eq:RS}
 \delta_{RS}=-\frac{1}{2}q_i\sigma_i q_k \sigma_k
\end{equation}
changing from CDR to DRED. Here $i$ denotes the emitter, $k$ denotes the spectator. $q_{i,k}$ are the fractional charges
of emitter and spectator respectively, and $\sigma_{i,k}$ are the sign factors as defined in Eq.~(\ref{eq:signfactor}).
However there is only a contribution in case 
of massless emitters. Soft singularities are independent of the scheme and therefore a scheme dependence only arises from
collinear splittings. As these are finite in the massive case there is no transition term for massive emitters.
Eventually the full transition term is obtained by summing over all possible emitter-spectator pairs.

\subsection{Examples}
\label{susect:examples}
\subsubsection{$Z\to \nu \bar{\nu}$}
\begin{figure}[t]
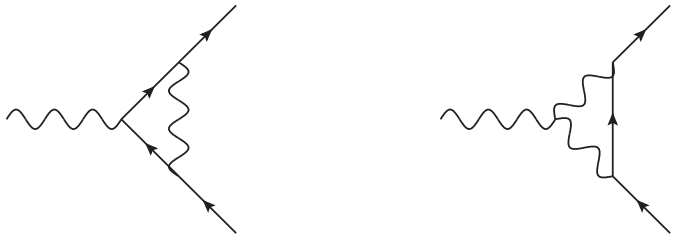

\centering
 \includegraphics[width=0.2\textwidth]{figs/nua.epsi}$\qquad\qquad\qquad$%
 \includegraphics[width=0.2\textwidth]{figs/nub.epsi}%
\caption{One loop electroweak corrections to $V \to f \bar{f}'$. The diagram on the left gives rise 
to the $\mathcal{V}_a$ contributions, while the $\mathcal{V}_b$ terms come from the diagram on the right.}
\label{fig:nuanub}
\end{figure}
As a first simple example of the applications
of the formulas of the previous subsections we consider the decay $Z\to \nu \bar{\nu}$. At the LO the corresponding matrix 
element is: 
\begin{equation}
\mathcal{M}_0 = i e g^{-}_{\nu} \bar{u}(p_1) \gamma^{ \mu } \omega_{-} v(p_2) \varepsilon_{\mu}(p_1 + p_2),
\label{eq:zlo}
\end{equation}
where $g^{-}_f = (I_f - s_W^2 Q_f)/s_W c_W$, $\omega_{\pm}=(1 \pm \gamma^5 )/2$ and 
$\varepsilon_{\mu}$ is the polarization vector for the on-shell external $Z$ boson.

The two classes of diagrams contributing to the one loop electroweak corrections to $Z\to \nu \bar{\nu}$ 
are shown in Fig.~\ref{fig:nuanub}. The unrenormalized $\mathcal{O}(\alpha)$ corrections to the $Z$ decay 
in neutrinos read:
\begin{eqnarray}
\mathcal{M}_{\rm Virt.} =  \mathcal{M}_0 \frac{ \alpha }{ 4 \pi } \Big\{
                          &  \frac{1}{4 s_W^2 c_W^2} & \mathcal{V}_a (0,M_Z^2,0,M_Z,0,0) \nonumber \\
&  + \frac{s_W^2 - \frac{1}{2}}{ s_W^2} & \mathcal{V}_a (0,M_Z^2,0,M_W,0,0)  \nonumber \\
&  + \frac{c_W^2}{ s_W^2} & \mathcal{V}_b (0,M_Z^2,0,0,M_W,M_W)  \Big\} ,
\label{eq:znlo}
\end{eqnarray}
where we used the notation of Ref.~\cite{Denner:1991kt} for the vertex functions $\mathcal{V}_{a(b)}$. It is worth 
noting that Eq.~(\ref{eq:znlo}) does not depend on the regularization scheme. What is scheme dependent is 
the expression of the $\mathcal{V}_{a(b)}$ functions. When performing the calculation in CDR we find:  
\begin{eqnarray}
& \mathcal{V}_a^{\rm CDR} (0,s,0,m,0,0)  =& -2s C_0 (0,s,0,m,0,0) \Big( 1+\frac{m^2}{s} \Big)^2 -
                                              B_0(s,0  ,0) \Big(3 + 2 \frac{m^2}{s} \Big) \nonumber \\
                                        & & + 2 B_0(0,m,0) \Big(2 + \frac{m^2}{s} \Big) -2 \\
& \mathcal{V}_b^{\rm CDR} (0,s,0,0,m_1,m_2)  =& 2 \Big( m_1^2 + m_2^2 + \frac{m_1^2 m_2^2}{s} \Big) C_0 (0,s,0,0,m_1,m_2) \nonumber \\ 
                                        & &  -B_0(s,m_1,m_2) \Big(1 + \frac{m_1^2+m_2^2}{s} \Big) \nonumber \\
                                        & & +  \Big(2 + \frac{m_1^2}{s} \Big) B_0(0,0,m_1^2)  
                                            +  \Big(2 + \frac{m_2^2}{s} \Big) B_0(0,0,m_2^2) ,
\label{eq:nuab}
\end{eqnarray}
(where $B_0$ and $C_0$ are the usual Passarino-Veltman two and three point functions, 
respectively~\cite{'tHooft:1978xw,Passarino:1978jh,Bardin:1999ak}), 
while the DRED results are related to CDR ones as follows:
\begin{eqnarray}
& \mathcal{V}_a^{\rm DRED} (0,s,0,m,0,0)      =& \mathcal{V}_a^{\rm CDR} (0,s,0,m,0,0)  +1 \nonumber \\
& \mathcal{V}_b^{\rm DRED} (0,s,0,0,m_1,m_2)  =& \mathcal{V}_b^{\rm CDR} (0,s,0,0,m_1,m_2)+1 .
\label{eq:nuabconv}
\end{eqnarray}

The $\mathcal{O}(\alpha)$ corrections to $Z\to \nu \bar{\nu}$ coming from renormalization read:
\begin{equation}
\mathcal{M}_{\rm CT   } =  \mathcal{M}_0 \Big\{ \frac{1}{2} \delta Z_Z + \delta Z_{ \nu }^L 
                           + \delta Z_e + \Big( \frac{s_W^2}{c_W^2} -1\Big) \frac{\delta s_W}{ s_W} \Big\} .
\label{eq:zct}
\end{equation}
Using Eq.~(\ref{eq:nuabconv}) and the scheme conversion rules for the counterterms in eqs.~(\ref{eq:convBOS}) 
and (\ref{eq:convlept}), it is easy to show that 
$( \frac{ \mathcal{M}_{\rm Virt.} + \mathcal{M}_{\rm CT}}{\mathcal{M}_0} )^{\rm CDR } =$
$( \frac{ \mathcal{M}_{\rm Virt.} + \mathcal{M}_{\rm CT}}{\mathcal{M}_0} )^{\rm DRED}$. 

Note that Eq.~(\ref{eq:zct}) contains large logarithms of of the light fermion masses, 
that can be removed by using $\alpha(M_Z^2)$ as input parameter in Eq.~(\ref{eq:zlo}) 
and performing the corresponding finite renormalization 
$\delta Z^e |_{ \alpha( M_Z^2 ) } \to \delta Z^e |_{ \alpha(0)} - \frac{1}{2} \Delta \alpha(M_Z^2) $, 
where 
\begin{equation}
\Delta \alpha(Q^2) = \Pi^{AA}_{f \in \, {\rm light } }(0) - {\rm Re} \big( \Pi^{AA}_{f \in \, {\rm light } }(Q^2) \big),  
\qquad  \Pi^{AA}_{f \in \, {\rm light } }(Q^2) = \frac{ \Sigma^{AA}_{f \in \, {\rm light } } (Q^2)}{Q^2}, 
\label{eq:deltaalpha}
\end{equation}
and $\Sigma^{AA}_{f \in \, {\rm light }}$ represents the contribution of light fermion loops 
to the photon self energy (which has the same expression both in DRED and in CDR). Once the singular 
terms in the $m_q \to 0$ limit have been subtracted, the remaining dependence on the light quark 
masses can be safely neglected.

\subsubsection{$W^{+} \to u \bar{d}$}
With the notation of Eq.~(\ref{eq:zlo}), the matrix element for the process $W^{+} \to u \bar{d}$ is:
\begin{equation}
\mathcal{M}_0 = i \frac{e}{\sqrt{2} s_W} \bar{u}(p_1) \gamma^{ \mu } \omega_{-} v(p_2) \varepsilon_{\mu}(p_1 + p_2). 
\label{eq:wlo}
\end{equation}

The one loop electroweak corrections to $W^{+} \to u \bar{d}$ are given by the two classes of diagrams 
in Fig.~\ref{fig:nuanub}. At variance with the $Z \to \nu \bar{\nu}$ case, in this case also photon loops 
contribute to the corrections. We basically follow the computation of Ref.~\cite{Denner:1991kt}, with the important 
difference that we use dimensional regularization for the IR singularities coming from photon loops. These IR 
singularities are cancelled by the QED dipole subtraction terms described in subsection~\ref{susec:elinop}. 

The unrenormalized virtual one loop matrix element is: 
\begin{eqnarray}
\mathcal{M}_{\rm Virt.} =  \mathcal{M}_0 \frac{ \alpha }{ 4 \pi } \Big\{
      &  Q_u Q_d         & \mathcal{V}_a (0,M_W^2,0,0,0,0) \nonumber \\
      &+ g_u^{-} g_d^{-} & \mathcal{V}_a (0,M_W^2,0,M_Z,0,0)  \nonumber \\
      &+ Q_u             & \mathcal{V}_b (0,M_W^2,0,0, 0 , M_W)\nonumber \\  
      &- Q_d             & \mathcal{V}_b (0,M_W^2,0,0,M_W,   0)\nonumber \\
      &+ g_u^{-} \frac{c_W}{s_W} & \mathcal{V}_b (0,M_W^2,0,0,M_Z, M_W) \nonumber \\  
      &- g_d^{-} \frac{c_W}{s_W} & \mathcal{V}_b (0,M_W^2,0,0,M_W, M_Z) \Big\} ,
\label{eq:wnlo}
\end{eqnarray}
while the contribution coming from renormalization can be written as:
\begin{equation}
\mathcal{M}_{\rm CT   } =  \mathcal{M}_0  \Big\{ \frac{1}{2} \delta Z_W + \frac{1}{2}\delta Z_u^L + \frac{1}{2}\delta Z_d^L    
                           + \delta Z_e - \frac{\delta s_W}{ s_W} \Big\}.
\label{eq:wct}
\end{equation}

As in the previous example, eqs.~(\ref{eq:wnlo}-\ref{eq:wct}) are completely general, while the expression 
of the vertex functions and the counterterms depend on the regularization scheme. If we define $\delta^{\rm Virt.}$ via 
the relation: 
\begin{equation}
\label{eq:wdvirtdef}
2 {\rm Re } \Big[ \Big( \mathcal{M}_{\rm Virt.} + \mathcal{M}_{\rm CT   } \Big)^{*} \mathcal{M}_0 \Big] 
= \frac{ \alpha }{ 2 \pi } \delta^{\rm Virt.} | \mathcal{M}_0 |^{2},
\end{equation}
from eqs.~(\ref{eq:nuabconv}), (\ref{eq:convUquark2}-\ref{eq:convDquark2}) and~(\ref{eq:convBOS}), we 
find that the one loop virtual electroweak corrections have the following regularization scheme dependence:
\begin{equation}
\label{eq:wudschemedep}
\delta^{\rm Virt.}_{ \rm DRED} -\delta^{\rm Virt.}_{ \rm CDR} = \frac{1}{2}Q_u^2+\frac{1}{2}Q_d^2 \;.
\end{equation}  
Also in this case, the scheme 
dependence of the one loop electroweak corrections is canceled by the corresponding scheme dependence
of the QED integrated dipoles that are obtained from Eq.~(\ref{eq:RS}) by summing over all emitter and spectator pairs. 
As a result, the sum of the renormalized one loop virtual corrections and the infrared subtraction terms is regularization 
scheme independent. As in the previous example, the dependence on the light quark masses in Eq.~(\ref{eq:wct}) 
can be removed by using $\alpha(M_Z^2)$ or $\alpha_{G_{\mu}}$ as input parameters.

\subsubsection{$u \bar{u} \to c \bar{c}$}
As a last simple example we consider the one loop electroweak corrections to the $\mathcal{O}(\alpha_S)$ 
process $u \bar{u} \to c \bar{c}$. As the $\mathcal{O}(\alpha)$ corrections are helicity-dependent, it is 
convenient to write the LO matrix element as:
\begin{equation}
\mathcal{M}_0 = \frac{i g_S^2 T^a_{ij} T^b_{kl}}{(p_u + p_{\bar{u}})^2} 
\sum_{\lambda_i,\lambda_f = \pm } 
\bar{u}(p_c)         \gamma^{ \mu } \omega_{\lambda_f} v(p_{\bar{c}})
\bar{v}(p_{\bar{u}}) \gamma_{ \mu } \omega_{\lambda_i} u(p_u) = 
\sum_{\lambda_i,\lambda_f = \pm } \mathcal{M}_0 (\lambda_i,\lambda_f) .
\label{eq:louucc}
\end{equation}

\begin{figure}[t]
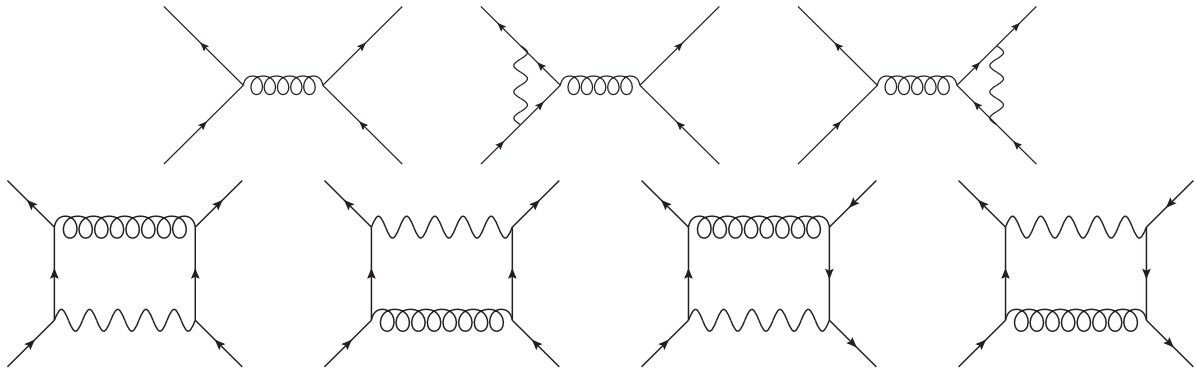

\centering
 \includegraphics[width=0.2\textwidth]{figs/uuccborn.epsi}$\qquad$%
 \includegraphics[width=0.2\textwidth]{figs/uuccvi.epsi}$\qquad$%
 \includegraphics[width=0.2\textwidth]{figs/uuccvf.epsi}\\
 \includegraphics[width=0.2\textwidth]{figs/uuccbox1.epsi}$\qquad$%
 \includegraphics[width=0.2\textwidth]{figs/uuccbox2.epsi}$\qquad$%
 \includegraphics[width=0.2\textwidth]{figs/uuccbox3.epsi}$\qquad$%
 \includegraphics[width=0.2\textwidth]{figs/uuccbox4.epsi}%
\caption{LO and virtual NLO electroweak diagrams for the process $u \bar{u} \to c \bar{c}$.}
\label{fig:uucc}
\end{figure}

Fig.~\ref{fig:uucc} represents the diagrams contributing to the NLO electroweak corrections to the process 
$u \bar{u} \to c \bar{c}$. As in the previous example, we regularize dimensionally the  
QED-like IR singularities coming from the photon loops and also the QCD-like IR singularities associated 
with the gluon exchange between the fermionic currents in the initial and in the final states. 
For a given helicity configuration, the virtual electroweak corrections read:
\begin{equation}
\mathcal{M}_{\rm Virt.} (\lambda_i,\lambda_f) =  \mathcal{M}_0 (\lambda_i,\lambda_f) \frac{ \alpha }{ 4 \pi } \Big\{ 
\delta^{\rm Vertex}(\lambda_i) + \delta^{\rm Vertex}(\lambda_f) +\delta^{\rm Box} (\lambda_i,\lambda_f) \Big\} ,
\label{eq:nlouucc}
\end{equation}
with 
\begin{eqnarray}
\delta^{\rm Vertex}(\lambda) = \Big\{
     &  (g_u^{\lambda})^2 & \mathcal{V}_a (0,(p_u + p_{\bar{u}})^2,0,M_Z,0,0) \nonumber \\
     & +Q_u^2             & \mathcal{V}_a (0,(p_u + p_{\bar{u}})^2,0,0  ,0,0)  \nonumber \\
     &  +\delta_{\lambda,-} \frac{1}{4 s_W^2} & \mathcal{V}_a (0,(p_u + p_{\bar{u}})^2,0,M_W,0,0) \Big\} .
\label{eq:vertuucc}
\end{eqnarray}

The counterterm contribution is:
\begin{equation}
\mathcal{M}_{\rm CT   } (\lambda_i,\lambda_f) 
=  \mathcal{M}_0 (\lambda_i,\lambda_f)  \Big\{ \delta Z_u^{\lambda_i} + \delta Z_u^{\lambda_f} \Big\},
\label{eq:ctuucc}
\end{equation}
where $\delta Z_u^{+}=\delta Z_u^{R}$, $\delta Z_u^{-}=\delta Z_u^{L}$ and $\delta Z_u^{L(R)}=\delta Z_c^{L(R)}$. 

It is worth noting that the box contribution in Eq.~(\ref{eq:nlouucc}) does not depend on the   
regularization scheme (CDR/DRED) and does not give rise to rational terms neither of UV 
type (since it is UV finite) nor of IR type~\cite{Bredenstein:2008zb}. 
As a consequence, using eqs.~(\ref{eq:nuabconv}) 
in~(\ref{eq:vertuucc}) together with the scheme conversion formulas for the counterterms~(\ref{eq:convUquark2}) 
we can show that for each of the  helicity configurations the one loop electroweak corrections depend 
on the regularization scheme according to:
\begin{equation}
\label{eq:uuccschemedep}
\delta^{\rm Virt.}_{ \rm DRED}(\lambda_i,\lambda_f) -\delta^{\rm Virt.}_{ \rm CDR}(\lambda_i,\lambda_f) =
 2 Q_u^2,
\end{equation}  
where $\delta^{\rm Virt.}(\lambda_i,\lambda_f)$ corresponds to Eq.~(\ref{eq:wdvirtdef}) with the 
replacement $\mathcal{M} \to \mathcal{M}(\lambda_i,\lambda_f)$ (of course Eq.~(\ref{eq:uuccschemedep}) 
also holds at the level of unpolarized matrix elements). Also in this case, the scheme dependence 
of the one loop electroweak corrections is cancelled by the corresponding scheme dependence of the 
QED integrated dipoles.

\section{Real emission and subtraction procedures}
\label{real}
Conceptually the calculation of the real emission processes and the appropriate subtraction
terms for the cancellation of QED infrared singularities in numerical simulations is very similar to the case of
QCD calculations. Many concepts and methods that were originally developed for QCD calculations can easily
be adapted to electroweak computations. 
The additional radiation of a photon is a source of infrared singularities in regions of phase space where
the photon becomes either soft or collinear to other charged massless particles. For sufficiently inclusive 
observables, the KLN theorem~\cite{Kinoshita:1962ur,Lee:1964is} 
guarantees the cancellation of infrared singularities from final state particles against 
their counterparts from the virtual contributions. Collinear initial state singularities have to be either    
absorbed into the parton distribution functions, or these singularities have to be regulated using the physical
mass of the incoming particles. The corresponding logarithms of the masses are a potential source of large corrections.

In the context of one loop electroweak calculations, the real contributions are basically defined as the 
real QED corrections to the process under consideration. Nevertheless, also the radiation of an 
additional $W/Z$ boson would contribute at the same order in perturbation theory. In the literature, 
the extra emission of additional $W$ and $Z$ bosons usually is not included in the computation 
of real $\cal{O}( \alpha )$ corrections for two reasons. First of all the gauge bosons decay, so 
that in principle they lead to final states which are not degenerate with the LO ones. The second 
reason is that real purely weak corrections are always finite: in fact, even in the Sudakov regime, where 
the corrections are dominated by the Sudakov logarithms which are the IR limit of virtual weak corrections, 
the gauge boson masses act as physical IR regulators. However, in those kinematical regions where the Sudakov 
corrections become large, the effect of the partially compensating radiation of real gauge bosons may lead 
to significant positive contributions. In Refs.~\cite{Ciafaloni:2006qu,Ciafaloni:2003xf,Ciafaloni:2001vu,Ciafaloni:2001vt,
Ciafaloni:2000rp,Ciafaloni:2000df,Bell:2010gi,Stirling:2012ak,Manohar:2014vxa,Frixione:2014qaa,Frixione:2015zaa} 
the issue of real weak boson emission 
has been addressed in analogy with QED or QCD: 
all the diagrams obtained from the LO ones with the emission of an additional $W$ or $Z$ 
boson are considered as real corrections, the additional gauge boson is produced on-shell and integrated 
over the full phase space. The result is a significant cancellation between real and virtual corrections, 
which however may be incomplete due to the incomplete average on the isospin of the initial state 
particles~\footnote{For example, in $e^{+}e^{-}$ collisions the cancellation would require also the processes with 
$e^{+} \nu_e$, $e^{-} \overline{\nu_e}$ and $\nu_e \overline{\nu_e}$ as initial states, while in 
$pp$ collisions the cancellation would take place if $u$ and $d$ type quarks were weighted by the 
same PDFs.} 
(Bloch Nordsieck violating effects~\cite{Ciafaloni:2001vt,Ciafaloni:2000rp,Ciafaloni:2000df}). 
Real weak boson emission processes have been considered from a more phenomenological point of view in 
Refs.~\cite{Baur:2006sn,Bell:2010gi,Bern:2012vx,Chiesa:2013yma}: the additional gauge bosons decay and they are 
included in the real corrections only when the final states are degenerate with the LO ones. 
Also in this case, the cancellation between real 
and virtual weak corrections is only partial and strongly dependent on the event 
selection under consideration. Recently, two different implementations of multiple 
weak boson emission processes in the context of parton showers have been presented 
in Refs.~\cite{Christiansen:2014kba} and~\cite{Krauss:2014yaa}. 
%The inclusion of real weak boson 
%emission effect would however lead to an at least partial cancellation of the 
%electroweak Sudakov logarithms that are the leading part of the virtual EW corrections at high 
%energies~\cite{Ciafaloni:2000df,Ciafaloni:2000rp,Ciafaloni:2001vt,Ciafaloni:2006qu,
%Bell:2010gi,Stirling:2012ak,Baur:2006sn,Bern:2012vx,Chiesa:2013yma}. 

From the theoretical point of view the real emission of a photon can be handled in the same way as it is 
done in QCD, namely by defining jets as infrared safe observables, and a jet may or may not contain a photon.
A clean way of doing that is the 'democratic clustering' approach~\cite{Glover:1993xc} where all the partonic final 
states (hadrons and photons) are treated on equal footing.
In that sense the photon is then treated fully inclusively and the results are infrared safe.
For many final states however the presence of an additional photon can experimentally be tagged and therefore 
be distinguished from the original Born process. For instance a muon being collinear to a photon can be disentangled
as they end up in different parts of the detector. From the theoretical point of view, this implies that the 
photon is not treated fully inclusively but parts of the collinear singularity is cut away. This leads to the 
problem that the collinear singularity in the virtual corrections is not fully canceled by the corresponding  
one in the real emission. As the muon mass acts as physical cut-off for the collinear singularities, the only 
effect of the miscancellation is a large QED correction. However, the incomplete cancellation would be an issue 
in the case of the photon radiation off a massless particle.
There are two options to handle this problem. The first starts from the observation that a photon can 
either be produced in the hard scattering, i.e. it can be treated perturbatively, or it occurs much later
at lower energies, when hadrons are created and decay into more stable hadrons. This happens in an energy 
regime that is non-perturbative and which can be described by the photon fragmentation function~\cite{Glover:1993xc}.
The photon fragmentation function is an observable that has to be measured experimentally, 
similarly to the parton distribution functions.
And in the same way as initial state collinear singularities can be absorbed into the parton distribution functions,
the part of the collinear singularity that is not canceled by the virtual corrections can be absorbed into the photon
fragmentation function. The second option to deal with this incomplete cancellation is to (also experimentally)
only cut away finite pieces such that the cancellation of the singularity remains unspoiled. 
This can be achieved by reducing the allowed amount of photonic energy within the jet as a function of 
the R-separation between photon and tagged 
particle such that the photon energy vanishes in the limit of collinearity~\cite{Frixione:1998jh}.
We note that this problems also occur in fixed order QCD calculations where photons are present.

Additional complications arise from the fact that for a general process one cannot distinguish between
QCD corrections and electroweak corrections but all contributions at a given order in $\alpha$ and
$\alpha_s$ have to be taken into account. For the real emission this includes contributions from a gluon
exchange between the interference of a QCD Born diagram with an electroweak Born diagram. 
It is worth to stress that these contributions are therefore not the absolute squared of an amplitude but 
just the interference term between two different types of amplitudes. They have different properties
concerning their color -but also their kinematical- structure compared to standard tree-level like contributions. 
These terms 
can be handled with standard methods from QCD and we refer to section~\ref{wjj} for a concrete example,
nevertheless their automated generation is in general not trivial (but it is within the reach of the tools 
dealing with automated calculations). 

\section{Electroweak corrections to $W^{+}$ plus two jets}
\label{wjj}
In the last part of this review we show a computation that has
been constructed in a quasi automated way interfacing the {\sc{GoSam}} and
{\sc{MadDipole}} programs. We calculate the electroweak NLO corrections to the production
of a $W^{+}$ boson in association with two jets,
which has also been calculated in Ref.~\cite{Kallweit:2014xda}. 
In the present calculation we treat the $W$-boson as a stable particle. Although this might be of limited interest from a purely
phenomenological point of view, the decay of the $W$ 
does not add any fundamental complication to the calculation which
is not already present in the approach of a stable $W$-boson. 
%(of course, in the case of an off-shell $W$ production, the calculation should be rephrased in the context 
%of the complex mass scheme 
%and the IR subtraction procedure should be modified in order to deal with the presence of 
Of course, in the case of an off-shell $W$ production, the calculation should be rephrased in the context 
of the complex mass scheme; moreover,  
the IR subtraction procedure might be modified in order to have an efficient 
integration of real matrix elements plus local counterterms also in the presence of 
resonances~\cite{Basso:2015gca,Jezo:2015aia}. 
The complete NLO computation should include the decay of the gauge 
boson together with the contributions of photon induced processes. For 
simplicity, we neglect both these contributions although they could be 
computed in the same framework of our calculation. By the way, these 
contributions will not affect significantly the inclusive results, 
but in particular the photon initiated processes can be important in 
the tails of distributions: indeed, they could well be the subject 
of a dedicated phenomenological study~\footnote{Photon induced contributions 
have been included in Ref.~\cite{Kallweit:2014xda}.}.
For all these reasons, we stress that in this review the process 
$W+2$~jets serves as a proof of concept where one faces many subtleties 
of an electroweak computation while at the same time minimizing the 
necessary computational efforts. 
%This process serves us as a proof of concept where one faces
%all the subtleties of an electroweak computation while at the same time minimizing the necessary computational efforts.
In fact, in many aspects this process is identical to the calculation of the dijet process, which has been presented
in Ref.~\cite{Dittmaier:2012kx}.\\
\subsection{Computational setup}
\begin{figure}[t]
\centering
 \includegraphics[width=0.7\textwidth]{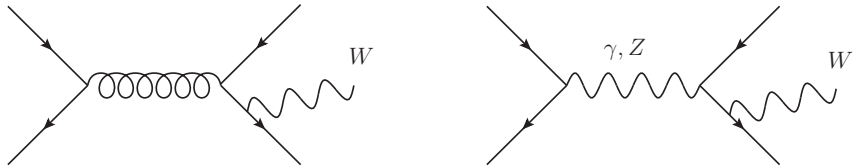}
\caption{Example diagrams for a QCD Born (left diagram) and an electroweak Born (right diagram). At leading order only
the left diagrams are taken into account.}
\label{fig:born}
\end{figure}

The starting point is a Born contribution of the order ${\cal O}(\alpha_s^2 \alpha)$. 
At this order in the coupling constants two classes of diagrams contribute to the Born process: 
the ones with two external quarks (with two additional gluons either radiated off the quarks or 
produced via gluon splitting) and the ones with four quarks involving a gluon exchange between the 
two fermionic currents. In principle there would be also a Born contribution of the order
${\cal O}(\alpha^3)$ involving only diagrams with four quarks where the internal gluon is 
replaced by a photon or a massive gauge boson. However, for the NLO EW corrections to $W+$2~jets, 
one can safely neglect the ${\cal O}(\alpha^3)$ underlying Born: it would lead to ${\cal O}(\alpha^4)$ 
contributions which will be much smaller than the ${\cal O}(\alpha_s^2 \alpha^2)$  
ones due to the different sizes of the coupling constants. 
Two sample diagrams are depicted in Fig.~\ref{fig:born}.\\

The fact that the $W$-boson is treated as a stable particle raises an additional complication at one loop. 
In fact, the building blocks of the calculation involve diagrams with internal massive propagators that 
can give rise to singularities when the momentum flowing in these propagators approaches the on-shell 
limit. On one hand these singularities could be avoided by means of the complex mass scheme, which would 
imply that the on-shell prescription for the process $W+2$~jets should be abandoned. On the other hand, 
a naive inclusion of the widths in the massive propagators would lead to inconsistencies because of 
the different treatment of the external and the internal $W$ bosons (in particular, the soft singularities 
corresponding to the $W-\gamma$~splitting would be turned into logarithms of the widths). However, 
in the process $W+2$~jets at ${\cal O}(\alpha_s^2 \alpha^2)$ 
these potentially problematic diagrams are interfered with a QCD Born that does not involve  internal massive 
particles. 
This means that these terms do not give rise to physical resonances, however 
they are only integrable as principal value integrals. For a numerical Monte Carlo integration 
this is however problematic as it would simply try to integrate over a singular region. 

As discussed in Ref.~\cite{Kallweit:2014xda}, one possible solution is to introduce a regulator width~$\Gamma$ in 
those pseudo-resonant diagrams that are not associated with IR singularities, so that the remaining 
diagrams are free of pseudo-resonances. 
Though this procedure technically breaks gauge invariance, the violation of gauge invariance 
is of order $\Gamma/M$. 
For the one loop contribution we adopt a different method that is based on a local K-factor. 
For each phase space point we require that the finite part of the one loop contribution normalized 
to the underlying QCD Born (i.e. a Born that is free of resonances) has 
to be smaller than a certain value. The advantage of this approach is that one does not have to introduce 
a width for only specific diagrams. 
Nevertheless it can be related to the method of using a regulator width: in fact, a large K-factor is obtained 
if the internal particle is close to being on-shell and the K-factor effectively cuts the contribution 
off if the particle is too close to being on-shell. One can roughly estimate the width corresponding 
to a given K-factor by comparing the massive propagator to a massless one, requiring that
\begin{equation}
\label{eq:k}
 \frac{p^2}{p^2-m^2} < K\;,
\end{equation}
and solving Eq.~(\ref{eq:k}) for $p$ for a given value of $K$. This way one obtains the minimally 
allowed distance from the mass $m$. 
It is worth noting that the procedure in Eq.~(\ref{eq:k}) is only an approximation, as in general 
many other aspects enter in the K-factor. 
As a default we used a value of $1000$ for the K-factor, but we varied its value from one hundred to 
one thousand and found that the results for the different values of the K-factor agree within their 
statistical errors. To further reduce numerical instabilities we generated 
the diagrams with a Higgs exchange separately and used an integration channel that resembles 
exactly the peak structure of these diagrams. This may serve as an additional strategy to improve numerical 
convergence for parton level results: loop diagrams from which one expects different
peak structures can be generated separately and then integrated with a dedicated parametrisation of the phase space.

For the subtraction terms consisting of a gluon exchange between an EW Born and a QCD Born as illustrated in 
Fig.~\ref{fig:insertion} one can in principle use the same method, which is however not very convenient, 
as one has to perform this check for each subtraction term with its remapped kinematics. And it also 
requires to calculate the matrix element of the QCD amplitude. Therefore we adopt the method of 
Ref.~\cite{Kallweit:2014xda} and used a regulating width for the internal $W$- and $Z$-bosons. We varied
this with between $0.1$ and $1$ GeV and found agreement between the results within their statistical uncertainties.

\subsubsection{Virtual corrections}
\begin{figure}[t]
\centering
 \includegraphics[width=0.8\textwidth]{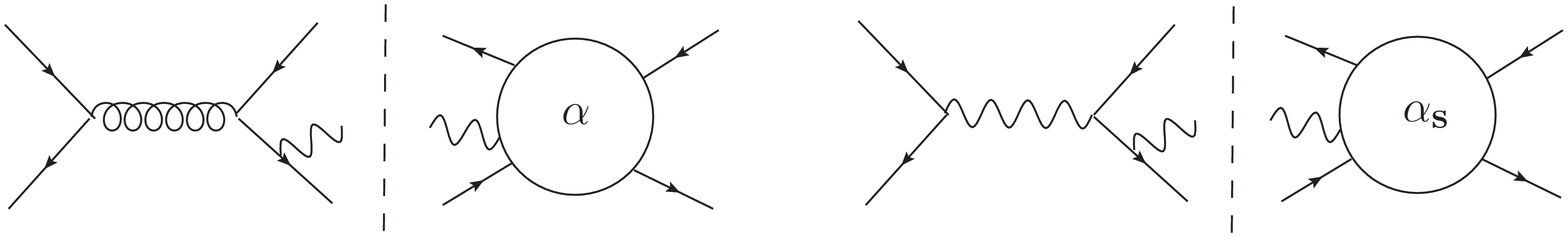}
 \includegraphics[width=0.6\textwidth]{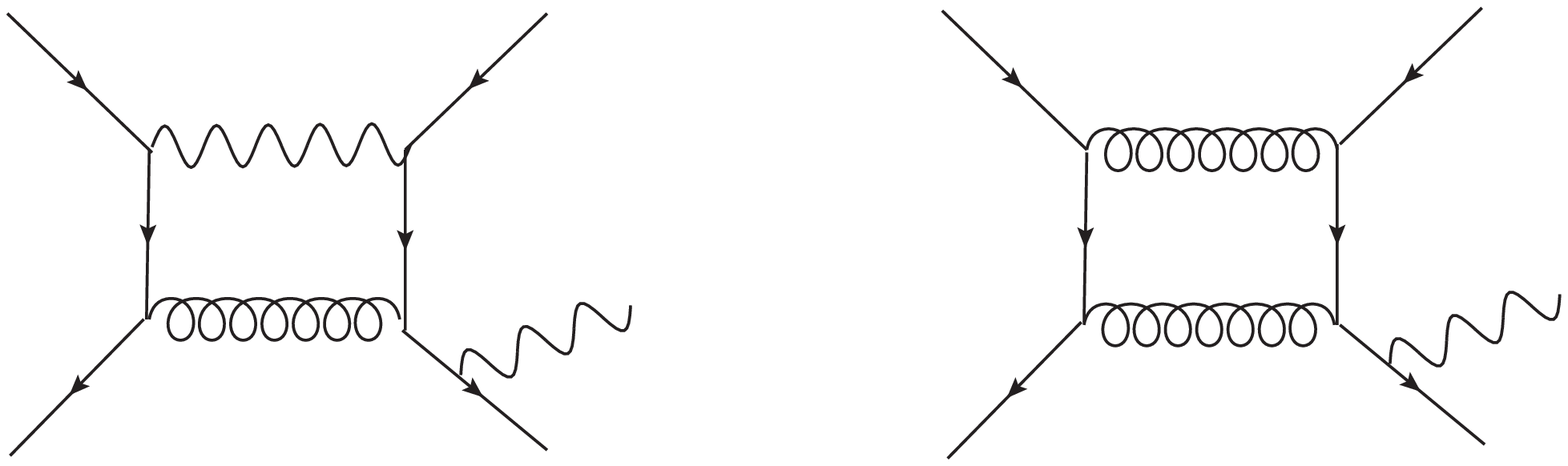}
\caption{Different virtual contributions. In the upper row the left plot denotes the interference between a QCD tree-level diagram with an electroweak
one-loop diagram, the right plot displays the interference between an electroweak tree-level diagram and a QCD one-loop diagram.
As the two pieces contribute at the same order both have to be taken into account.
The lower row shows an example diagram for each of the two types of one-loop amplitudes.}
\label{fig:virtual}
\end{figure}

For this process we have to distinguish two different types of virtual corrections. Starting from the QCD like born structure
we have to calculate the one-loop electroweak corrections leading to a contribution of ${\cal O}(\alpha_s^2\alpha^2)$. 
However this order in perturbation theory can also be achieved by interfering a loop amplitude of ${\cal O}(\alpha_s^2\alpha^{1/2})$
with a pure electroweak tree-level amplitude of ${\cal O}(\alpha^{3/2})$. This contribution can be understood as an interference
term between an electroweak tree-level amplitude with a QCD one-loop amplitude originating from the QCD born. The two different
contributions are illustrated in Fig.~\ref{fig:virtual}. The lower two plots show an example diagram for the two cases. 
The left diagram contains a loop with both a gluon and a photon (or a weak gauge boson) in the loop. 
This is a good example that the distinction
between QCD corrections and electroweak corrections does not make sense here, as this diagram can be regarded to be a QCD 
correction to an electroweak Born as well as an electroweak correction to a QCD born. This also means that these kinds of
diagrams might contain singularities from both interactions, QCD and electroweak. The loop on the right hand is a pure QCD 
correction and therefore only contains QCD singularities.\\

Simply speaking this situation always arises in the case where at tree-level the desired final state can be obtained in two different
ways, either with a QCD Born or with an electroweak Born. The two contributions can be calculated independently, however, 
only the sum of the two yields a physically meaningful result and one has to sum up all the terms that contribute to a given
order in perturbation theory at next-to-leading order, both in $\alpha_s$ and $\alpha$.\\
There is one subtle difference between the two contributions. In the first case one starts from a QCD born amplitude with
a given order in $\alpha$ and $\alpha_s$ and the one-loop amplitude just adds another factor of $\alpha$  while keeping the 
original order in 
$\alpha_s$. This is also the situation one encounters performing QCD corrections and this is in some sense the standard way.
The second contribution however is different. Starting from an electroweak born amplitude, the corresponding one-loop amplitude
is two powers higher in $\alpha_s$ but also one power lower in $\alpha$. In view of automating electroweak calculations and combining
different tools, in particular one-loop providers (OLP) and Monte Carlo tools (MC), this is a special case that is not included
in the currently used standard interface \cite{Binoth:2010xt,Alioli:2013nda}. Therefore an appropriate extension of the
existing interface would be highly desirable.\\

In this calculation the virtual amplitudes are generated with {\sc{GoSam}} \cite{Cullen:2011ac,Cullen:2014yla}. 
The {\sc{GoSam}} framework is based on an algebraic generation of $D$-dimensional integrands using a Feynman
diagrammatic approach, employing
{\sc{QGRAF}}~\cite{Nogueira:1991ex} and
{\sc{FORM}}~\cite{Vermaseren:2000nd,Kuipers:2012rf} for the diagram generation, and
{\sc{Spinney}}~\cite{Cullen:2010jv}, {\sc{Haggies}}~\cite{Reiter:2009ts} and
{\sc{FORM}} to write an optimized Fortran output. For the reduction of the tensor integrals
we used {\sc{Ninja}}~\cite{Mastrolia:2012bu,vanDeurzen:2013saa,Peraro:2014cba}, a package for the integrand reduction via Laurent expansion.
One can however also use different reduction
techniques such as integrand reduction via the OPP method~\cite{Ossola:2006us,Mastrolia:2008jb,Ossola:2008xq} as implemented in
{\sc{Samurai}}~\cite{Mastrolia:2010nb} or one can use methods of tensor reduction as the ones contained in 
{\sc{Golem95}}~\cite{Heinrich:2010ax,Binoth:2008uq,Cullen:2011kv,Guillet:2013msa}.
The remaining scalar integrals have been evaluated using {\sc{OneLoop}}~\cite{vanHameren:2010cp}.
All these methods and tools have originally been developed for the case of QCD one-loop corrections. However for none of them
it matters whether one uses them for QCD corrections or for electroweak corrections. The generation of the diagrams, and
the reduction of the tensor integrals or the integrand reduction respectively rely on general properties that do not 
depend on the type of the interaction.\\
As {\sc{GoSam}} allows to specify
the orders of tree-level amplitude and one-loop amplitude separately, i.e. does not rely on the fact that the one-loop amplitude is
just one order higher in the coupling constant, it is straightforward to obtain the two different virtual contributions.
We have extended the {\sc{GoSam}} framework to incorporate the electroweak counter terms as well as the electroweak
 infrared insertion operator. 
\begin{figure}[t]
\centering
 \includegraphics[width=1\textwidth]{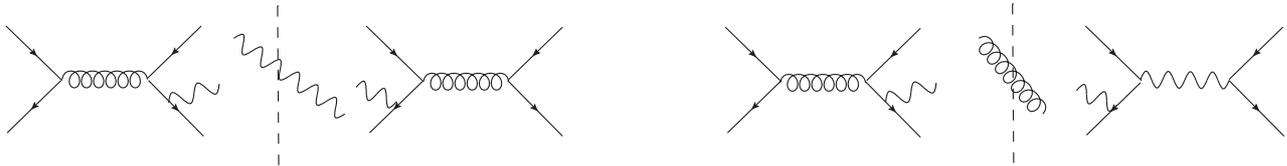}
\caption{Schematical representation of the two different kinds of insertion operators, representing QED singularities (l.h.s.) and
QCD singularities (r.h.s.). This also illustrates the two different types of real emission contributions.}
\label{fig:insertion}
\end{figure}

The advantage of the inclusion of the insertion operator is twofold. In {\sc{GoSam}} 
the agreement of the poles is the first level of the trigger for numerical instabilities.
The numerical structure of the insertion operator is very simple and it is proportional to the Born. However on the side of the one-loop
amplitude this simple results is obtained in a highly non-trivial way. Numerical instabilities can be detected as deviations from the
correct coefficient of the pole terms. This can then either be used to trigger some kind of rescue system or it can be used to
make quantitative estimation on the accuracy of the obtained result.
As the pole cancellation only works for the renormalized amplitude this is also an important check for the electroweak counter terms,
which are new in the {\sc{GoSam}} framework. 
Especially for these types of processes, where one has different virtual contributions, 
the check of this pole cancellation is very important as 
the cancellation only takes place when combining electroweak and QCD one loop contributions.\\
We should stress that the formulas for the insertion operator in Eq.~(\ref{eq:I}) are only valid for 
contributions from the l.h.s of Fig.~\ref{fig:virtual}.
They are obtained by photon insertion between the underlying QCD amplitudes, i.e. the insertion 
operator only covers the QED singularities of these
diagrams. It does not cover any singularity of QCD origin which are contained in the diagrams of the r.h.s. of Fig.~\ref{fig:virtual}.
Therefore, in general one does not find pole cancellation between the two terms in Fig.~\ref{fig:virtual} and the QED insertion 
operator.
The singularities from contributions stemming from the r.h.s of Fig.~\ref{fig:virtual} are of purely QCD origin. This means that we need
 to apply a QCD insertion operator. In contrary to normal QCD corrections, the born-like matrix element is given by the interference term
between a QCD Born amplitude and an electroweak Born amplitude with an additional gluon insertion that modifies the color structure of
the interference term (as it does in normal QCD). The singularities from this insertion operator will then cancel against the QCD 
singularities from one-loop contributions of both l.h.s. and r.h.s diagrams in Fig.~\ref{fig:virtual}. 
The two different types of insertion operators are illustrated in Fig.~\ref{fig:insertion}.\\
For the production of a $W$-boson in association with jets
the number of processes that 
need to be generated can be reduced to:
\begin{flalign}
\label{subprocs}
 \bar{u}u &\to W^+ \bar{u}d \nonumber \\
 u \bar{d}&\to W^+ c\bar{c}\nonumber\\
 u \bar{d}&\to W^+ s\bar{s}\nonumber\\
 u \bar{d}&\to W^+ gg\nonumber\\
 \bar{u}u &\to W^+ \bar{c}s\nonumber\\
 \bar{d}d &\to W^+ \bar{c}s\nonumber\\
 u\bar{d} &\to W^+ d\bar{d} \, .
\end{flalign}
Note that for the process $u \bar{d}\to W^+ gg$ there exists no underlying electroweak born, hence this process does not have
a QCD one-loop contribution.\\
We have checked the pole cancellation mentioned above for all the independent subprocesses. All other processes can be obtained
by crossing. 
Besides the renormalization of the electroweak sector this type of process also requires the renormalization of QCD singularities.
They arise from the QCD loop diagrams that are interfered with an electroweak born process. The renormalization procedure itself
is straightforward and identical to a conventional QCD calculation, however the interpretation is a bit different.
The underlying born that needs to be renormalized is an 
interference term between the QCD born and the electroweak born rather than the absolute square of an amplitude at a given order
in $\alpha_s$.

\subsubsection{Regularization scheme dependence}
In the {\sc{GoSam}} framework dimensional reduction is the more appealing approach 
as one deals with a simplified $D$-dimensional Dirac algebra 
and does not need to take care about additional renormalization terms for $\gamma_5$. This typically leads to
more compact expressions and hence is the method of choice. 
It is worth mentioning that for the given process, the scheme dependence for the QCD loops is proportional to the interference
term between the QCD Born and the electroweak Born, with an additional modified color factor. For subprocesses that only have
one color structure the scheme dependence vanishes. This is however not a general feature and does not hold for more than one
color factor.

\subsubsection{Real emission}
The real emission contributions contain processes based on an underlying QCD Born with an additional photon. 
We neglect additional radiation
of massive electroweak gauge bosons. As discussed above, a second contribution stems from the radiation of an additional gluon
from an underlying electroweak Born amplitude, interfered with a QCD Born amplitude. 
This contribution is necessary to cancel the QCD like singularities from the virtual corrections. All real emission
contributions and corresponding subtraction terms for both QCD and QED radiation have been generated with 
{\sc{MadGraph}} 4 \cite{Stelzer:1994ta,Alwall:2007st}
and {\sc{MadDipole}} \cite{Frederix:2008hu,Frederix:2010cj,Gehrmann:2010ry}, 
containing subtraction terms based on the dipole formalism as
developed in \cite{Catani:1996vz,Dittmaier:1999mb}. The QED version of {\sc{MadDipole}} has 
been extended to cover the photon radiation off the
$W$-boson. The integration over tree-level and real emission phase space has been 
performed using {\sc{MadEvent}}~\cite{Maltoni:2002qb}.
The inclusion of the virtual contributions has been obtained using the BLHA interface, 
which has been adapted to be usable for electroweak corrections.

\subsubsection{Parameters}
As a consistent set of input parameters we use $M_W$, $M_Z$ and $G_{\mu}$ with
\begin{equation}
 M_W= 80.398 \,\text{GeV}, \quad M_Z= 91.1876\, \text{GeV}, \quad G_{\mu}=1.1663787\cdot 10^{-5} \, \text{GeV}^{-2},
\end{equation}
while $\alpha$ is a derived quantity:
\begin{equation}
 \alpha_{G_{ \mu }}=\frac{\sqrt{2}G_{\mu}M_W^2}{\pi}\left(1-\frac{M_W^2}{M_Z^2}\right).
\end{equation}
The $G_{ \mu }$ input scheme avoids large logarithms involving light fermion masses originating from the running 
of $\alpha$ 
from the Thomson limit to the electroweak scale~\cite{Denner:1991kt,Dittmaier:2012kx}. As these logarithms 
are effectively included in the definition of the coupling constant, they have to be subtracted from 
electric charge renormalization counterterm with the following 
replacement~\cite{Denner:1991kt,Dittmaier:2001ay,Dittmaier:2012kx}:
\begin{equation}
\delta Z^e |_{G_{ \mu }} = \delta Z^e |_{ \alpha(0)} - \frac{1}{2} \Delta r,
\end{equation}
where $\delta Z^e |_{ \alpha(0)}$ is defined in Eq.~(\ref{eq:rencond3}) while $\Delta r$ summarizes the 
radiative corrections to muon decay~\cite{Sirlin:1980nh,Marciano:1980pb,Sirlin:1981yz} 
apart from the QED corrections which coincide with those of the Fermi model.
Due to the real emission of the photon, which happens at $q^2=0$ one power of the coupling constant $\alpha$
is taken to be as $\alpha=\alpha(0)=1/137.035999074$. 
Moreover, the one loop renormalization is a function 
of the Higgs and the top quark mass. In our numerical studies we used the values:
\begin{alignat}{6}
\label{eq:inpother}
m_e & =  510.99892 \; {\rm KeV},& \qquad   m_{\mu}  & =105.658369 \;  {\rm MeV},& \qquad  m_{\tau} &=1.77699\;  {\rm GeV}, & \nonumber  \\
m_t & =  172.9 \; {\rm GeV},&  \qquad M_H        &= 120 \;  {\rm GeV}.& & & &
\end{alignat}
As the $W$ is treated as a stable particle in this calculation, no complex mass scheme is applied here.

Renormalization and factorization scale are both set to $M_W$, the center of mass energy is chosen to be 14~TeV.
As this is a calculation which is leading order in QCD we use the MSTW2008lo pdf set.
The CKM matrix has been chosen to be a diagonal unit matrix.
In order to set cuts in an infrared safe way we use the following approach of democratic clustering:
 all particles, including the $W$-boson are passed through the jet algorithm and clustered according to an anti-kt algorithm
 \cite{Cacciari:2005hq,Cacciari:2008gp} contained in the {\sc{Fastjet}} package~\cite{Cacciari:2011ma}.
 In a second step we however require that the $W$-boson as well as two jets containing
a QCD parton are tagged. This ensures that we are also infrared safe from a QCD point of view.
The resulting jets need to fulfill the following basic cuts:
\begin{equation}
 p_T > 20 \, \text{GeV}, \quad |\eta|<4.4, \quad \Delta R > 0.4\,.
\end{equation}
This is in particular also true for the $W$-boson.

\begin{figure}[t]
\centering
 \includegraphics[width=0.49\textwidth]{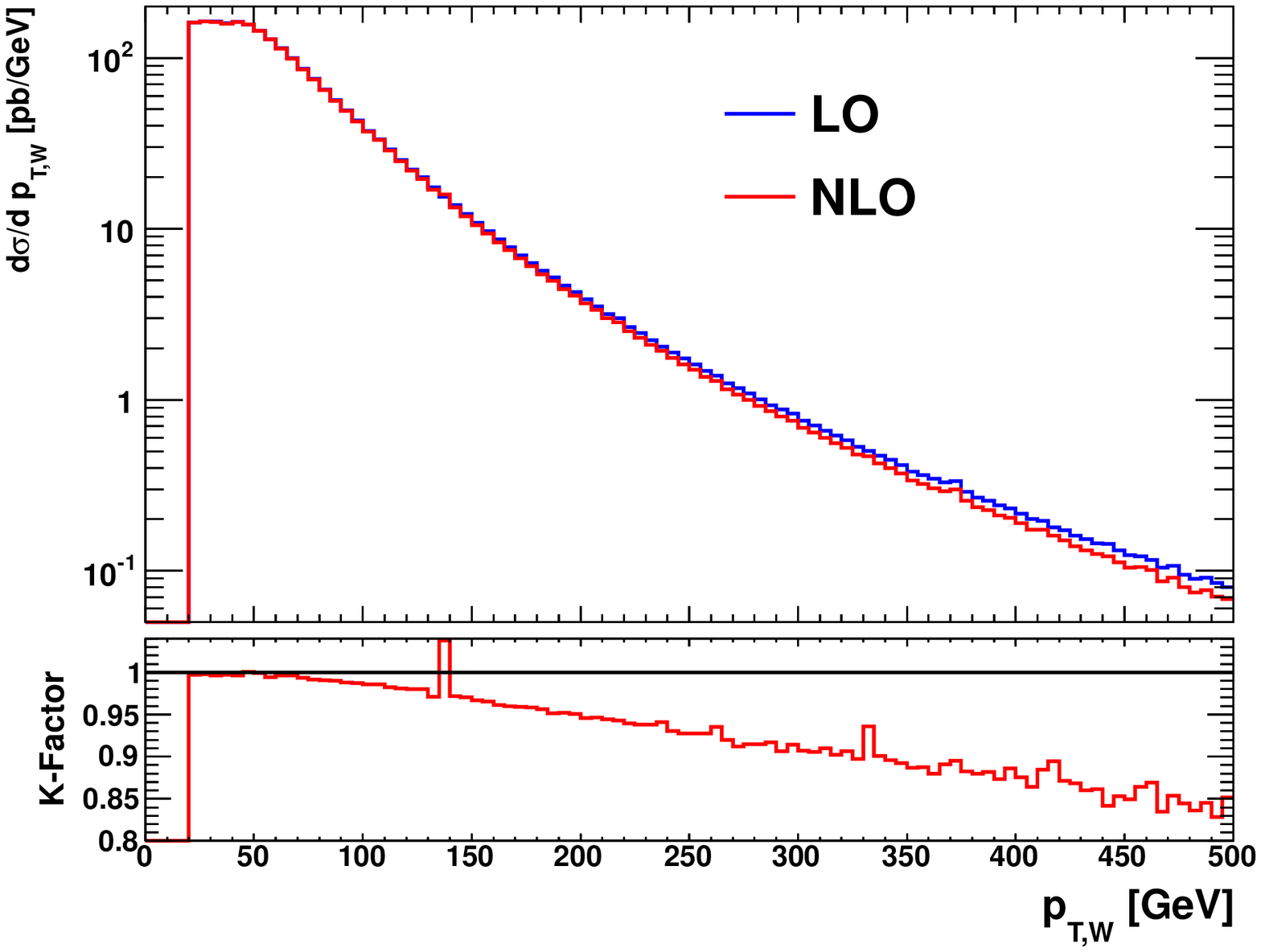}
 \includegraphics[width=0.49\textwidth]{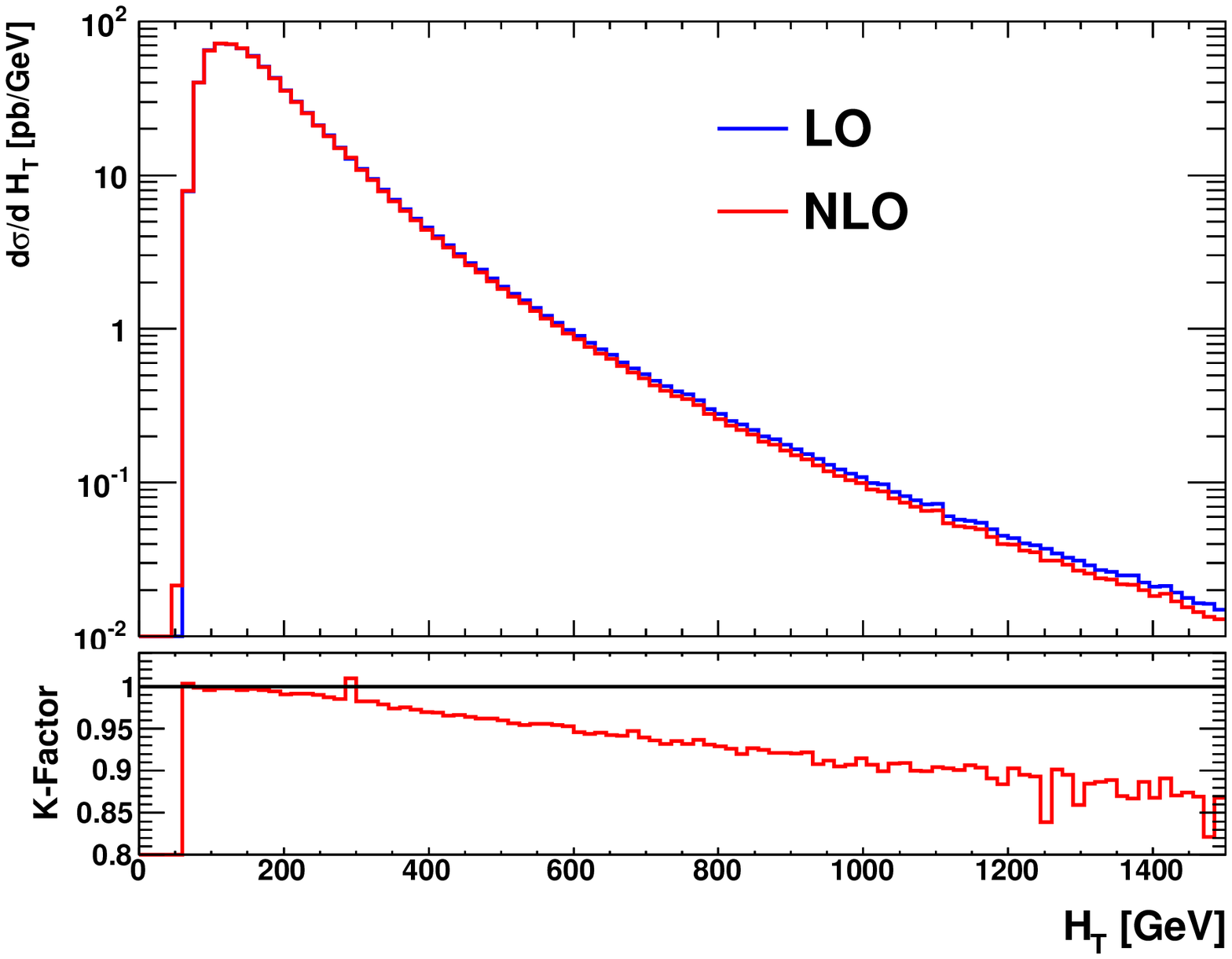}
\caption{Differential distribution of the transverse momentum of the $W$-boson (left plot) and the $H_T$ distribution (right plot).}
\label{fig:ptw}
\end{figure}

\subsection{Numerical Results}
With the set of parameters stated above we find a total cross section at leading order of
\begin{equation}
 \label{eq:lores}
 \sigma_{tot,LO}= 10873 \pm 6 \text{pb}\;.
\end{equation}
As expected, the electroweak NLO corrections only have a small impact on the total cross section. Their contribution is
negative and we find a total cross section of
\begin{equation}
 \label{eq:nlores}
 \sigma_{tot,NLO} = 10789 \pm 6 \text{pb}\;.
\end{equation}
Therefore we find a correction of $\sim 1\%$ to the total cross section. The errors in both eqs.~(\ref{eq:lores}) 
and~(\ref{eq:nlores}) represent the 1~$\sigma$ Monte Carlo integration error. 

More interesting than the total cross section are however differential distributions as one expects deviations stemming
from contributions of large logarithms of the $W$ mass (see e.g. \cite{Kuhn:2007qc}). 
In Fig.~\ref{fig:ptw} we show the transverse momentum of the $W$ boson. Although the impact in the low energy region is negligible
we do not obtain a constant K-factor but we see a substantial decrease of the cross section in the region where the transverse
momentum of the $W$ is large compared to its mass. In this region electroweak corrections typically supersede QCD corrections leading
to a $\sim 20\%$ correction in the region of $500$ GeV. 
A similar behavior is observed for the $H_T$ distribution, also leading to a $\sim 20\%$ in the high energy tail of the distribution.
This behavior stresses the importance of electroweak corrections at the LHC
in the context of searches for new physics, as one would typically search for new physics in the high energy limit. For observables
that are sensitive to Sudakov logarithms, like the transverse momenta or the $H_T$ distribution, the size of the electroweak
corrections grows beyond the uncertainty from the QCD corrections in the tails of the distributions. 

\begin{figure}[t]
\centering
 \includegraphics[width=0.49\textwidth]{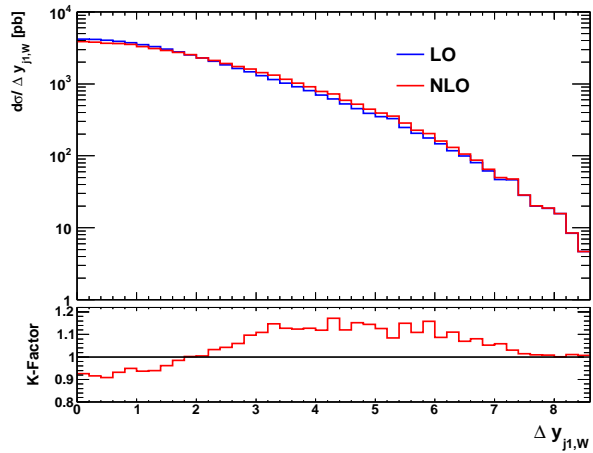}
\caption{Differential distribution for absolute value of the rapidity difference between the $W$ and the leading jet.}
\label{fig:deta}
\end{figure}

An interesting behavior can also be observed for the difference in rapidity between the $W$ and the jets 
$\Delta y_{j1,W}$. Fig.~\ref{fig:deta} shows
this observable for the leading jet, i.e. the jet with the highest transverse momentum. One observes a negative correction 
for small differences in rapidity, but a positive correction for large values of $\Delta y_{j1,W}$ 
which are however not relevant for the total cross section, 
as the contribution to the total cross section from this region is small.
The radiation of a particle with a high transverse momentum will more likely happen to take place in the central region, as one needs less
energy to get the same transverse momentum as in a more longitudinal direction. A $W$-boson with a high transverse momentum has to recoil
against the jets (or at least the leading jet). Therefore we expect the events with a high transverse momentum of the $W$ happening
for small differences in rapidity, which explains the negative next-to-leading order corrections in this region.
Turning this argument around we expect events with large differences in rapidity to come with smaller values of the transverse
momentum. This leaves more phase space to additional radiation which comes with a positive contribution and hence shifting the NLO
corrections to positive values.

\begin{figure}[t]
\centering
 \includegraphics[width=0.49\textwidth]{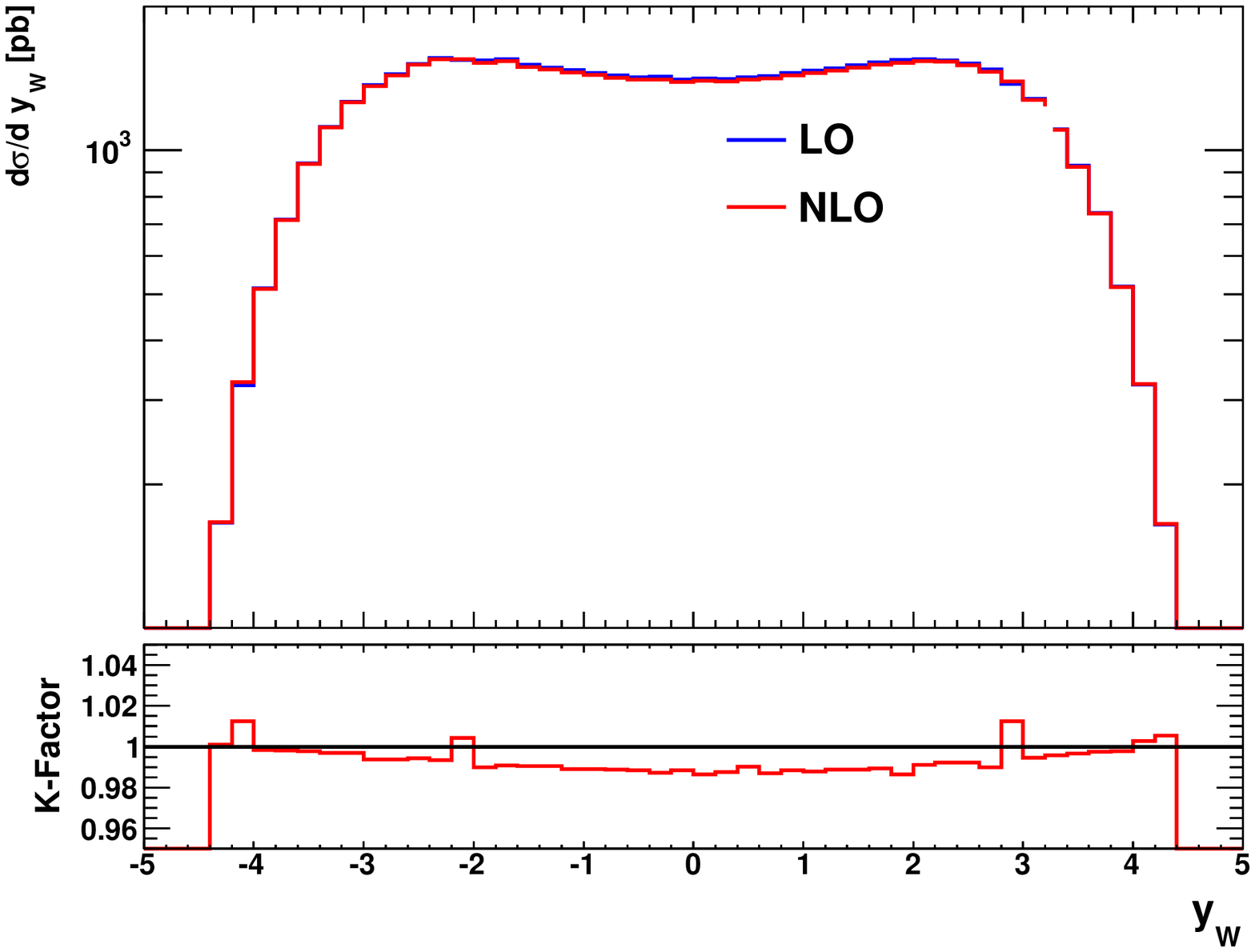}
 \includegraphics[width=0.49\textwidth]{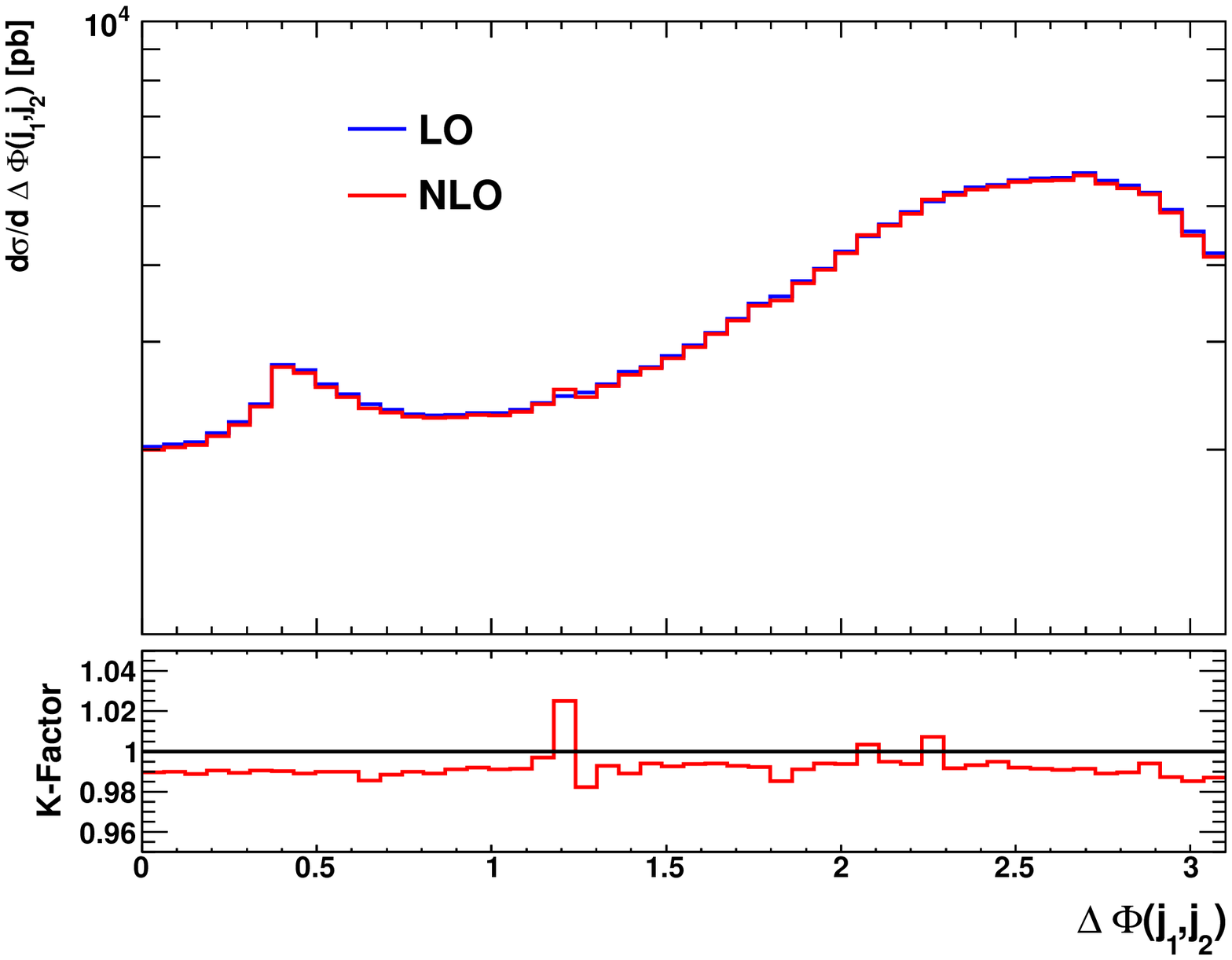}
\caption{Differential distribution of the rapidity of the $W$-boson (left plot) and the 
difference of the azimuthal angles of the two leading jets (right plot).}
\label{fig:ydphi}
\end{figure}

The same behavior is also found for the rapidity distribution of the $W$ which is shown in the left plot of Fig.~\ref{fig:ydphi},
although the rapidity distribution to a larger extents mixes events with different transverse momenta and therefore the
effect is much less pronounced. 
Generally speaking we do not expect visible effects of electroweak NLO corrections for observables where the events with 
high transverse momentum of the $W$ are evenly distributed 
as a function of the observable under consideration.  
One such example is shown in the right plot
of Fig.~\ref{fig:ydphi}, where the difference in the azimuthal angle between the two leading jets is displayed.
In that case we obtain a flat K-factor yielding a small negative correction at next-to-leading order.

\subsection{Virtual amplitudes for single phase space point}
In this section we give the results for all the ingredients that enter the virtual contribution for a single phase space point
reported in Table~I.
The regularization scheme is DRED, the couplings $g_s$ and $e$ are set to one for simplicity and all the results in the 
tables do not contain the overall factor $\alpha_{(s)}/2\pi$.
The other parameters are set as stated above. The notation is as follows:
Born-QCD denotes the tree-level matrix element squared 
that is of the order ${\cal O}(\alpha_s^2 \alpha)$, while
Born-EW denotes the tree-level matrix element squared of the order ${\cal O}(\alpha^3)$~\footnote{The Born-EW matrix elements
squared do not enter directly our numerical computation, nevertheless these numbers might be useful to fix completely 
the setup. Furthermore, the relative amplitudes have to be computed anyway to build essential interference terms.}.
The first line in the tables gives the contribution of the unrenormalized electroweak corrections, the second line the corresponding
counter terms. The third line denotes the unrenormalized contribution of QCD-loops interfered with the electroweak
tree-level amplitudes, followed by QCD renormalization terms. The last two lines then denote the contributions from 
the electroweak and QCD insertion operators.\\
\begin{center}
\begin{table}
\label{Phase}
\small{
\begin{tabular}{cccc}
 $E$ & $p_x$ & $p_y$ & $p_z$ \\
\toprule
250.0 & 0.0 & 0.0 & 250.0 \\
250.0 & 0.0 & 0.0 & -250.0 \\
111.48897463938037 & 56.330449672673069 & -24.583961957306506 & 46.781003797444882\\
207.38546864044983 & -167.21520572166636 & -118.01342209209260 & -33.475958310305167\\
181.12555672016978 & 110.88475604899321 & 142.59738404939910 & -13.305045487139747\\
\toprule
\end{tabular}
}
\caption{ Phase space point where all the relevant amplitudes of Eq.~(\ref{subprocs}) have been evaluated.}
\end{table}
\end{center}

\subsubsection{$\bar{u}u \to W^+ \bar{u}d$}
\label{uuwud}
{\centering {\small Born-QCD: $0.3331176339143377\rm{E}-03$ \qquad (Born-EW: $0.1159155705277286\rm{E}-02$)}}\\
{\small
\begin{tabular}[t]{|c|r|r|r|}
Process & \multicolumn{3}{|c|}{$\bar{u}u \to W^+ \bar{u}d$}  \\
        & $1/\epsilon^2$ & $1/\epsilon$ & finite \\
EW-Loop unren. &0.4515129528621876E-03 &0.5179344797596786E-02 &-0.4048755402433699E-02  \\
EW-Ren. & 0 &  -0.4682725341929206E-02 & -0.6484221636837220E-03 \\
QCD-Loop unren.  & 0.9326828685162938E-03 & -0.2347576391436467E-02 & 0.1872325091832032E-02\\
QCD-Ren. &0 & 0.1224146264927649E-02& 0.8851406118179909E-05 \\
I-EW &0.4811699156540433E-03 & -0.1471889938882662E-02& - \\
I-QCD & -0.1865365737032609E-02&  0.2098700609724008E-02& -\\
\hline
$\sum$ & -0.8E-16 & 0.1E-15& -0.2815845104231514E-02 \\
\end{tabular}
}
\hspace*{1cm}

 \subsubsection{$u\bar{d} \to W^+ c\bar{c}$}
 \label{udwcc}
 {\centering {\small Born-QCD: $0.1509105387580410\rm{E}-03$  \qquad (Born-EW:$0.3226028869165598\rm{E}-03$)}}\\
{\small
 \begin{tabular}[t]{|c|r|r|r|}
 Process & \multicolumn{3}{|c|}{$u\bar{d} \to W^+ c\bar{c}$}  \\
         & $1/\epsilon^2$ & $1/\epsilon$ & finite \\
 EW-Loop unren. &-0.2179818893171668E-03 & 0.2371604818917693E-02 & -0.7006797514364052E-03 \\
 EW-Ren. & 0 &  -0.2187116288266569E-02 & -0.3231845769861692E-03 \\ 
 QCD-Loop unren. &0 & -0.7013028153765194E-05 & 0.1644192724679622E-04  \\
 QCD-Ren. &0 & 0 & 0 \\
 I-EW &0.2179818893171704E-03 & -0.1915015588048693E-03& - \\
 I-QCD & 0 & 0.14026056307548736e-04 & -\\
 \hline
 $\sum$ & 0.4E-17 & 0.4E-16& -0.1007351745650022E-02 \\ 
 \end{tabular}
}
\hspace*{1cm}

 \subsubsection{$u\bar{d} \to W^+ s\bar{s}$}
 \label{udwss}
 {\centering {\small Born-QCD: $0.1509105387580410\rm{E}-03$ \qquad (Born-EW: $0.1683155172046103\rm{E}-03$)}}\\
{\small
 \begin{tabular}[t]{|c|r|r|r|}
 Process & \multicolumn{3}{|c|}{$u\bar{d} \to W^+ s\bar{s}$}  \\
         & $1/\epsilon^2$ & $1/\epsilon$ & finite \\
 EW-Loop unren. &-0.1173748634784783E-03 & 0.2359759858112650E-02 & -0.8847947525100161E-03 \\
 EW-Ren. & 0 &  -0.2201404574690460E-02 & -0.3267300831621827E-03 \\ 
 QCD-Loop unren. &0 & 0.1073376195468900E-04 & -0.8486634366399400E-04 \\
 QCD-Ren. &0 & 0 & 0 \\
 I-EW &0.1173748634784764E-03 & -0.1476215214675483E-03& - \\
 I-QCD & 0 & -0.2146752390936322e-04 & -\\
 \hline
 $\sum$ & -0.2E-17 & -0.3E-16& -0.1296320523810437E-02 \\ 
 \end{tabular}
}
\newpage

 \subsubsection{$u\bar{d} \to W^+ gg$}
 \label{udwgg}
 {\centering {\small Born-QCD: $0.4421271590046107\rm{E}-03$ \qquad (Born-EW: $0$)}}\\
{\small
 \begin{tabular}[t]{|c|r|r|r|}
 Process & \multicolumn{3}{|c|}{$u\bar{d} \to W^+ gg$}  \\
         & $1/\epsilon^2$ & $1/\epsilon$ & finite \\
 EW-Loop unren. &-0.2456261994470038E-03 & 0.5954936826411168E-02 & -0.4059861036705788E-02 \\
 EW-Ren. & 0 &  -0.5796026153495491E-02 & -0.6841797425946791E-03 \\ 
 QCD-Loop unren. &0 & 0 & 0 \\
 QCD-Ren. &0 & 0 & 0 \\
 I-EW &0.2456261994470059E-03 & -0.1589106729178175E-03& - \\
 I-QCD & 0 & 0 & -\\
 \hline
 $\sum$ & 0.2E-17 & -0.2E-14& -0.4218771709623606E-02 \\ 
 \end{tabular}
}
\hspace*{1cm}

 \subsubsection{$\bar{u}u \to W^+ \bar{c}s$}
 \label{uuwcs}
{ \centering {\small Born-QCD: $0.5471920731180749\rm{E}-04$ \qquad (Born-EW: $0.1560261025155393\rm{E}-03$)}}\\
{\small
 \begin{tabular}[t]{|c|r|r|r|}
 Process & \multicolumn{3}{|c|}{$\bar{u}u \to W^+ \bar{c}s$}  \\
         & $1/\epsilon^2$ & $1/\epsilon$ & finite \\
 EW-Loop unren. & -0.7903885500595288E-04& 0.1039825482424979E-02 & 0.6729212743821051E-03 \\
 EW-Ren. & 0 &  -0.7933882016361721E-03 & -0.1173430614024200E-03 \\ 
 QCD-Loop unren. &0 & 0.4658874285701936E-05 & -0.6180475632648689E-04 \\
 QCD-Ren. &0 & 0 & 0 \\
 I-EW &0.7903885500594414E-04 & -0.2417784065030774E-03& - \\
 I-QCD & 0 & -0.9317748571408931e-05 & -\\
 \hline
 $\sum$ & -0.9E-17 & 0.2E-16& 0.49379907590015959E-03 \\ 
 \end{tabular}
}
\newpage

 \subsubsection{$\bar{d}d \to W^+ \bar{c}s$}
 \label{ddwcs}
 {\centering {\small Born-QCD: $0.5471920731180750\rm{E}-04$ \qquad (Born-EW: $0.8665998896761033\rm{E}-03$)}}\\
{\small
 \begin{tabular}[t]{|c|r|r|r|}
 Process & \multicolumn{3}{|c|}{$\bar{d}d \to W^+ \bar{c}s$}  \\
         & $1/\epsilon^2$ & $1/\epsilon$ & finite \\
 EW-Loop unren. & -0.4255938346473633E-04& 0.8312276950839127E-03 & 0.1195260305972849E-02 \\
 EW-Ren. & 0 &  -0.7986225826314173E-03 & -0.1186419246527548E-03 \\ 
 QCD-Loop unren. &0 & -0.3536873862079221E-05 & -0.2058866127333486E-03 \\
 QCD-Ren. &0 & 0 & 0 \\
 I-EW & 0.4255938346473916E-04& -0.3614198631460137E-04& - \\
 I-QCD & 0 & 0.7073747724144110E-05 & -\\
 \hline
 $\sum$ & 0.2E-17 & -0.4E-16& 0.8707573878337068E-03 \\ 
 \end{tabular}
}
\hspace*{1cm}

 \subsubsection{$u\bar{d} \to W^+ d\bar{d}$}
 \label{udwdd}
 {\centering {\small Born-QCD: $0.8493852813137004\rm{E}-03$ \qquad (Born-EW: $0.1939842768622516\rm{E}-02$)}}\\
{\small
 \begin{tabular}[t]{|c|r|r|r|}
 Process & \multicolumn{3}{|c|}{$u\bar{d} \to W^+ d\bar{d}$}  \\
         & $1/\epsilon^2$ & $1/\epsilon$ & finite \\
 EW-Loop unren. & 0.1913723183724379E-02& 0.9838014169469896E-02 & -0.1220570133146914E-01 \\
 EW-Ren. & 0 &  -0.1197778947261055E-01 & -0.1665029533330621E-02 \\ 
 QCD-Loop unren. &0.2574356180301684E-02 & -0.6257504923291653E-02 & 0.1973036296675117E-02 \\
 QCD-Ren. &0 & 0.3378842486646013E-02 & 0.2443131831181329E-04 \\
 I-EW & 0.6606329965773225E-03&-0.8308733675698238E-03 & - \\
 I-QCD & -0.5148712360603448E-02 & 0.5849311107356306E-02 & -\\
 \hline
 $\sum$ & -0.6E-16 & 0.2E-15& -0.1187286557206206E-01 \\ 
 \end{tabular}
}

\section{Conclusions}
The recent start-up of the LHC's Run 2 will allow to test the Standard Model
through a set of new precision measurements. The search for new physics beyond 
the Standard Model is just started and new regions of phase space will be explored. 
The effect of the new physics might be large, but given the unique opportunity 
offered by the LHC physics program one has to look also for small deviations 
from the Standard Model. From the theoretical point of view this requires predictions 
based on precise and reliable computations whenever possible. This triggered 
the so called NLO revolution that was originally focused on providing NLO QCD corrections in 
an automated way. For many important processes like Higgs and top-quark production for instance
the NLO corrections were not precise enough to meet the requirements. This lead to an ongoing
active field of research namely pushing the limit further to NNLO QCD corrections.
On the other hand also electroweak corrections will play an important role 
in a large number of searches. 
Although their impact
on the total cross sections is typically rather small, they can lead to significant differences in distributions,
in particular in the high energy tails. However it is exactly the high energy limit that is sensitive
to new physics effects. Therefore, for a reliable comparison with LHC data from Run~2, 
the inclusion of electroweak higher order effects becomes inevitable for a large class of 
processes (also with large final state multiplicities). In this context, it is clear 
that fully automated tools for the computation of NLO electroweak corrections will play 
a crucial role at the LHC and at future hadron colliders.

We have presented an overview on the current status and recent development in the field of electroweak
NLO corrections for LHC physics. In particular we focused on the topic of the automation of 
$\mathcal{O}( \alpha )$ calculations, which is a very active and fast developing area of research. 
We have summarized the existing tools and the phenomenological predictions that have been 
obtained for LHC relevant processes. 
Furthermore we have discussed some of the general issues, related to the calculation 
of either the virtual one loop corrections or the real emission contributions, that 
have to be faced in order to develop an automated tool for the computation of NLO 
electroweak corrections starting from already available programs such as {\sc{GoSam}}  
(for the generation of one loop amplitudes) and {\sc{MadDipole}} (for the subtraction 
of infrared singularities). We discussed the regularization scheme dependence of 
the virtual one loop amplitudes, the renormalization counterterms and infrared 
subtraction terms (at least for the specific case of the CDR and DRED schemes) and 
we provided the corresponding scheme transition rules that are required in order 
to add up consistently these three building blocks of the calculation. We also 
pointed out the non trivial aspects of the infrared subtraction procedure in 
presence of QCD and QED-like infrared singularities. Lastly, as a non-trivial 
example, we have presented the full electroweak NLO corrections to the production 
of a $W$-boson in association with two jets within the {\sc{GoSam}}$+${\sc{MadDipole}} 
framework.

\section*{Acknowledgements}
We thank Stefan Dittmaier, Thomas Hahn, Wolfgang Hollik, Guido Montagna, Fulvio Piccinini and 
Sandro Uccirati for useful discussions. 
N.G. wants to thank the University of Naples and the University of Pavia for kind hospitality while parts of this
work were carried out. The work of F.T. was partially supported by the INFN Iniziativa Specifica PhenoLNF.
The work of M.C. was partially supported by the INFN Iniziativa Specifica QFTATCOLLIDERS.
The work of M.C. and F.T. were partially supported by the Italian Ministry of University and Research 
under the PRIN project 2010YJ2NYW.
\bibliographystyle{JHEP}
\bibliography{refs}
\end{document}